\documentclass[12pt]{iopart}
\usepackage{iopams} 
\usepackage{array}
\usepackage{graphicx} 
\usepackage{cite}
\usepackage{epstopdf}
\usepackage{etoolbox}
\usepackage{booktabs}
\usepackage{multirow}
\usepackage{slashbox}
\usepackage{hhline}
\usepackage{xcolor}

\renewcommand{\vec}[1]{\underline{#1}}
\renewcommand{\det}{\mathrm{det}}

\newcommand{\eqref}[1]{(\ref{#1})}
\newcommand{\hf}{\frac{1}{2}}

\newcommand{\binom}[2]{\left(\begin{array}{@{}c@{}}#1\\#2\end{array}\right)}
\newcommand{\euler}[2]{\left\langle\begin{array}{@{}c@{}}#1\\#2\end{array}\right\rangle}
\newcommand{\bra}[1]{\langle #1 |}
\newcommand{\ket}[1]{| #1\rangle}
\newcommand{\braket}[2]{\langle #1 | #2\rangle}
\newcommand{\qbinom}[2]{\left[ \begin{array} {@{}c@{}}#1 \\ #2\end{array} \right]_{\!q}}
\newcommand{\halfint}[1]{\lfloor \frac{#1}{2}\rfloor}

\makeatletter
\def\@mkboth#1#2{}
\newlength\appendixwidth
\preto\appendix{\addtocontents{toc}{\protect\patchl@section}}
\newcommand{\patchl@section}{%
	\settowidth{\appendixwidth}{\textbf{Appendix }}%
	\addtolength{\appendixwidth}{1.5em}%
	\patchcmd{\l@section}{1.5em}{\appendixwidth}{}{\ddt}%
}
\makeatother

\begin{document}
\maketitle
\topical{Combinatorial mappings of exclusion processes}

\author{Anthony J. Wood, Richard A. Blythe, Martin R. Evans}
\address{School of Physics and Astronomy, University of Edinburgh, Peter Guthrie Tait Road,	Edinburgh EH9 3FD}

\begin{abstract}
We review various combinatorial interpretations and mappings of stationary-state probabilities of the totally asymmetric, partially asymmetric and symmetric simple exclusion processes (TASEP, PASEP, SSEP respectively). In these steady states, the statistical weight of a configuration is determined from a matrix product, which can be written explicitly in terms of generalised ladder operators. This lends a natural association to the enumeration of random walks with certain properties.

Specifically, there is a one-to-many mapping of steady-state configurations to a larger state space of discrete paths, which themselves map to an even larger state space of number permutations. It is often the case that the configuration weights in the extended space are of a relatively simple form (e.g., a Boltzmann-like distribution). Meanwhile, various physical properties of the nonequilibrium steady state---such as the entropy---can be interpreted in terms of how this larger state space has been partitioned. 

These mappings sometimes allow physical results to be derived very simply, and conversely the physical approach allows some new combinatorial problems to be solved. This work brings together results and observations scattered in the combinatorics and statistical physics literature, and also presents new results. The review is pitched at statistical physicists who, though not professional combinatorialists, are competent and enthusiastic amateurs.
\end{abstract}

\tableofcontents

\section{Motivation: nonequilibrium stationary states and combinatorics}

A physical system is said to be in a nonequilibrium steady state (NESS) when the probability distribution over microstates is independent of time, but there are nevertheless nonzero currents of some quantity (such as energy, mass, or, more generally, probability). In the presence of these currents, the microstate probability distribution is not calculable by the conventional equilibrium statistical mechanics approach, where a partition function of Boltzmann weights would give access to all macroscopic observables. An open problem is to find an equivalent unified approach to solve a NESS.

This review focuses on a family of models with steady states that do permit exact solution: the asymmetric simple exclusion process (ASEP). These models, which involve hard-core particles hopping on a one-dimensional lattice, have endured for several reasons. First, they are motivated by physical phenomena, such as biophysical or vehicular transport \cite{schadschneider2010stochastic,Chou_2011}, which are of interest in their own right. Second, their analytical tractability has provided a number of fundamental insights into the behaviour of systems that are driven out of equilibrium, and how they contrast with equilibrium systems. For example, a major finding was that one-dimensional driven systems with short-range interactions can undergo phase transitions while their equilibrium counterparts can not \cite{lieb2013mathematical,evans2000phase}. Finally, the structure of the distribution over microstates that arises in the nonequilibrium setting is of interest from a more mathematical perspective.

It is this latter perspective that is the focus of this short review, with the particular aim of bringing together mathematical results that are scattered across the literature to the attention of statistical physicists. Our starting point is the matrix product solution of the ASEP, which was obtained by statistical physicists in the early 1990s \cite{derrida1992exact, derrida1993exactcorr,derrida1993exact,schuetz1993phase}. The first solution \cite{derrida1992exact} was obtained for special values of model parameters and was based on recursion relations between the statistical weight of configurations on lattices of different lengths. These recursion relations provide a clue that a deeper mathematical structure underlies this NESS. A major development was the introduction of \emph{matrix product} expressions for the stationary weights, wherein the recursion relations are replaced by algebraic rules that the matrices and vectors involved must obey \cite{derrida1993exact}. This approach allowed for physical quantities such as particle currents, density profiles and correlation functions, to be calculated more easily. Moreover, it also allowed solvable generalisations of the ASEP to be identified. The scope and application of the matrix product method is reviewed in detail in \cite{blythe2007nonequilibrium}.

Furthermore, the exact solution of the ASEP stationary state reveals, with increasing system size, number sequences that are ubiquitous within enumerative combinatorics. Specifically, in Section~\ref{sec:ASEP} below, we note the appearance of factorials (which count permutations), Catalan numbers (which count a variety of objects, including the number of legal combinations of nested brackets) and ballot numbers (which count subsets of nested bracket combinations). In other contexts, discussed in Section~\ref{sec:AlphaBeta1TASEP}, one also finds Narayana numbers (which count different subsets of nested brackets to ballot numbers) and Eulerian numbers (which count subsets of permutations). 

The main contribution of this review is to bring together and unify various results and observations in the literature which may shed light on the question: \emph{why should these combinatorial sequences arise in a nonequilibrium physics problem?}. At the broadest level, these sequences emerge from one-to-many mappings of configurations of particles in the ASEP to a larger set of objects (often, but not exclusively, paths on a lattice). For this we also introduce the idea of \emph{dominated paths}. This is a new way of illustrating the larger set of objects,  which offers an intuitive interpretation of the statistical weight of ASEP configurations, based on the paths they map to. We show how this is equivalent to mappings such as \emph{Motzkin} paths, that are already known in the literature.

This mapping fully describes the configuration space for when the hopping of the particles is entirely asymmetric. When the particles can hop either left or right, the set of objects vastly grows in size, to permutations of numbers. We outline a mapping from the combinatorial literature where the statistical weight of an ASEP configuration is found by enumerating permutations that follow certain rules. We then make an observation that these permutations can in fact be viewed as a mapping of the dominated paths discussed earlier. We show that this effectively interpolates between the totally asymmetric and partially asymmetric variants of the ASEP.

We also review some cases where dynamical rules can be defined on these larger state spaces that, when projected onto the ASEP configuration space, recover the appropriate ASEP dynamics. In particular, the dynamical transitions in the larger space may allow a bijection between the set of transition rates out of and into any configuration, implying a uniform distribution on the larger state space.

Finally we review generalisations to multispecies processes, where in certain cases the stationary state may be expressed as a matrix product. The combinatorial implications for the multispecies case are currently being explored and our aim is to point out emerging connections with queueing theory, integrable systems and associated algebraic combinatorics.

On a more practical level, we provide various examples throughout the review where physical results for the ASEP (such as correlation functions and density profiles) can be derived very efficiently using known results for these enumeration problems. Conversely the matrix product approach may help solve otherwise challenging combinatorial problems \cite{corteel2011matrix}. 

We begin the review by discussing the the totally asymmetric, single particle species exclusion process, which  turns out to be the easiest model to interpret in terms of combinatorics. We will introduce the mappings of this process, upon which the symmetric and partially asymmetric variants later generalise. We also discuss measures of entropy of these systems, and multispecies extensions. To begin, however, in the next section we give a summary of the matrix product formalism that solves these exclusion processes. The reader with some familiarity with these concepts may proceed to Section~\ref{sec:ThreeObservations} where we make some simple observations of the matrix product formalism that allude to this combinatorial structure, or to Section~\ref{sec:Remainder} where we set out the structure of this review in more detail.

\section{The asymmetric simple exclusion process}
\label{sec:ASEP}

\begin{figure}[!t]
	\centering
	\includegraphics{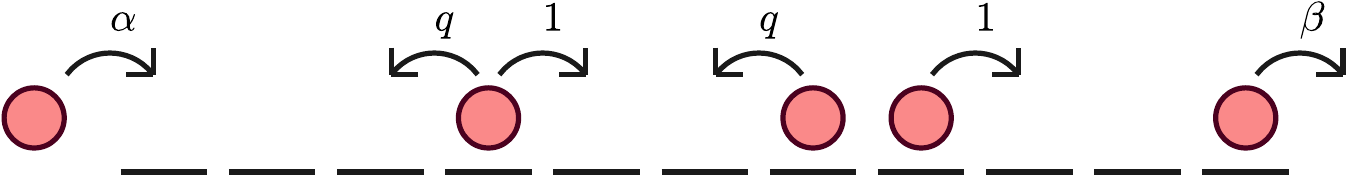}
	\caption{The exclusion process that constitutes the bulk of this review. Particles attempt to enter the system at rate $\alpha$, attempt to move to the right at rate $1$, to the left at rate $q$, and exit from the rightmost site at rate $\beta$. Particles may only move into free sites.}
	\label{fig:ASEPdiagram}
\end{figure}

The ASEP (Figure \ref{fig:ASEPdiagram}) is a stochastic open system of hopping particles. On a lattice of $N$ sites, particles can make unit steps left and right, at respective rates $q\geq0$ and $1$. They enter onto site $1$ from the left at rate $\alpha$, and leave from site $N$ at rate $\beta$. The system is referred to as the \emph{symmetric} simple exclusion process (SSEP) if $q = 1$, \emph{totally asymmetric} (TASEP) if $q=0$, and \emph{partially asymmetric} (PASEP) for all other $q$. The system is sometimes referred to as \emph{weakly asymmetric} (WASEP) if $q\to 1$ in a system-size-dependent way e.g.\ $q= 1-\mathcal{O}(N^{-1/2})$. In all variants the particles can only move onto vacant sites. In the long time limit, this system approaches a NESS, whereby the current of particles running left-to-right through the system stabilises.

This system is exactly solvable by a mathematical formalism known as the \emph{matrix product} approach, introduced in \cite{derrida1993exact}. The weight $\mathcal{W}$ of any configuration $\mathcal{C}$ of particles and vacant sites (or \emph{holes}) can be expressed as an ordered product of matrices, one per site, that represent the sequence of occupied and vacant sites along the lattice. As we now review, this approach permits the derivation of quantities such as the nonequilibrium partition function, average density profile and the steady-state current, as a function of $q$, $\alpha$, $\beta$. We note that the ASEP and matrix product solution can  be generalised to a five-parameter model with two additional parameters $\gamma$ and $\delta$ that correspond to exit of particles at the left and entry of particles at the right boundary \cite{derrida1993exact,USW2004,Uchiyama2008}. However in this review we restrict ourselves to the three parameter case.

\subsection{Overview of matrix product formalism of the ASEP}
\label{sec:matrixproducts}

Denote an ASEP configuration of size $N$ by $\mathcal{C} = (\tau_1, \tau_2, \dots \tau_N)$, where $\tau_i = 0$ if site $i$ is empty, and $1$ if it is occupied. The weight of $\mathcal{C}$ is an ordered product of matrices, reduced to a scalar by two boundary vectors $\bra{W}$, $\ket{V}$:
\begin{equation} \label{eq:MatrixProductAnsatz}
\mathcal{W}(\mathcal{C}) = \bra{W} \prod^N_{i=1} X_i(\tau_i) \ket{V} \;.
\end{equation}
Now, if we define matrices $X_i(1) = D$, $X_i(0) = E$ to represent occupied and vacant sites respectively, then the matrix product form \eqref{eq:MatrixProductAnsatz} for the weight of a configuration is a solution to the underlying steady-state master equation, given that these matrices and vectors have the following properties:
\begin{eqnarray}
DE &= qED + D+E \qquad\qquad \label{eq:ReductionRelations1} \\
D\ket{V} &= \frac{1}{\beta}\ket{V} \label{eq:ReductionRelations2} \\
\bra{W}E &= \frac{1}{\alpha}\bra{W} \label{eq:ReductionRelations3} \\
\braket{W}{V} &= 1 \;. \label{eq:ReductionRelations4}
\end{eqnarray}
Note that for the TASEP where $q=0$, relation \eqref{eq:ReductionRelations1} simplifies to
\begin{equation}
DE = D+E\;. \label{eq:ReductionRelations1TASEP}
\end{equation}

The first proof that the configurational weights (\ref{eq:MatrixProductAnsatz}) form a stationary solution of the master equation for the ASEP was given in \cite{derrida1993exact}. The \emph{quadratic algebra} (\ref{eq:ReductionRelations1}--\ref{eq:ReductionRelations3})
 can be used to reduce any string of matrices $\{D,E\}^N$ into strings that can be directly evaluated. As an example, the configuration $\mathcal{C} = (1, 0, 1)$ (a $3$-site ASEP) has weight 
\begin{eqnarray}
\label{eq:ExamplePASEPWeight}
W(1,0,1) &= \bra{W}DED\ket{V} \\
& = \bra{W}(D+E+qED)D\ket{V} \\
& = \bra{W}DD\ket{V}+ \bra{W}ED\ket{V}+q\bra{W}EDD\ket{V} \\
& = \frac{1}{\beta^2}+\frac{1}{\alpha\beta}+\frac{q}{\alpha\beta^2} \;,
\end{eqnarray}
where we have taken $\langle W | V \rangle =1$.
Here we see that repeated application of \eqref{eq:ReductionRelations1} generates a sum of matrix products. Once each term is in a \emph{normal-ordered} form with all $E$s to the left and $D$s to the right, Eqs.~\eqref{eq:ReductionRelations2} and \eqref{eq:ReductionRelations3} can be used to convert each term to a scalar. In the following we will associate each term with the weight of a configuration in an expanded space of configurations. If we think of the rates $\alpha$, $\beta$ and $q$ as each being the exponential of an energy-like quantity, then configurations in the expanded space can be thought of following a Boltzmann distribution with an appropriately-defined energy function. Then, each weight in the \emph{nonequilibrium} ensemble is given by the sum over weights in an \emph{equilibrium} ensemble. 

The partition function $Z_N$ of the nonequilibrium ensemble is the total weight of the $2^N$ distinct configurations of particles and holes. This can be written as
\begin{equation}
Z_N = \sum_{\mathcal{C}} \mathcal{W}(\mathcal{C}) = \bra{W}(D+E)^N\ket{V}
\end{equation}
since the expansion of $(D+E)^N$ yields all possible strings of $D$ and $E$ of length $N$. Explicit expressions for $Z_N$ are given below via Eqs.~\eqref{eq:SSEPAlphaBetaPartitionFunction} and  \eqref{eq:TASEPPartitionFunctionExpression}. The case for general $q$ is more complicated \cite{blythe2000exact}; for completeness we present this in \ref{sec:Zgen}.

\subsubsection{Phase diagram.}

\begin{figure}[!t]
	\centering
	\includegraphics[scale=0.8]{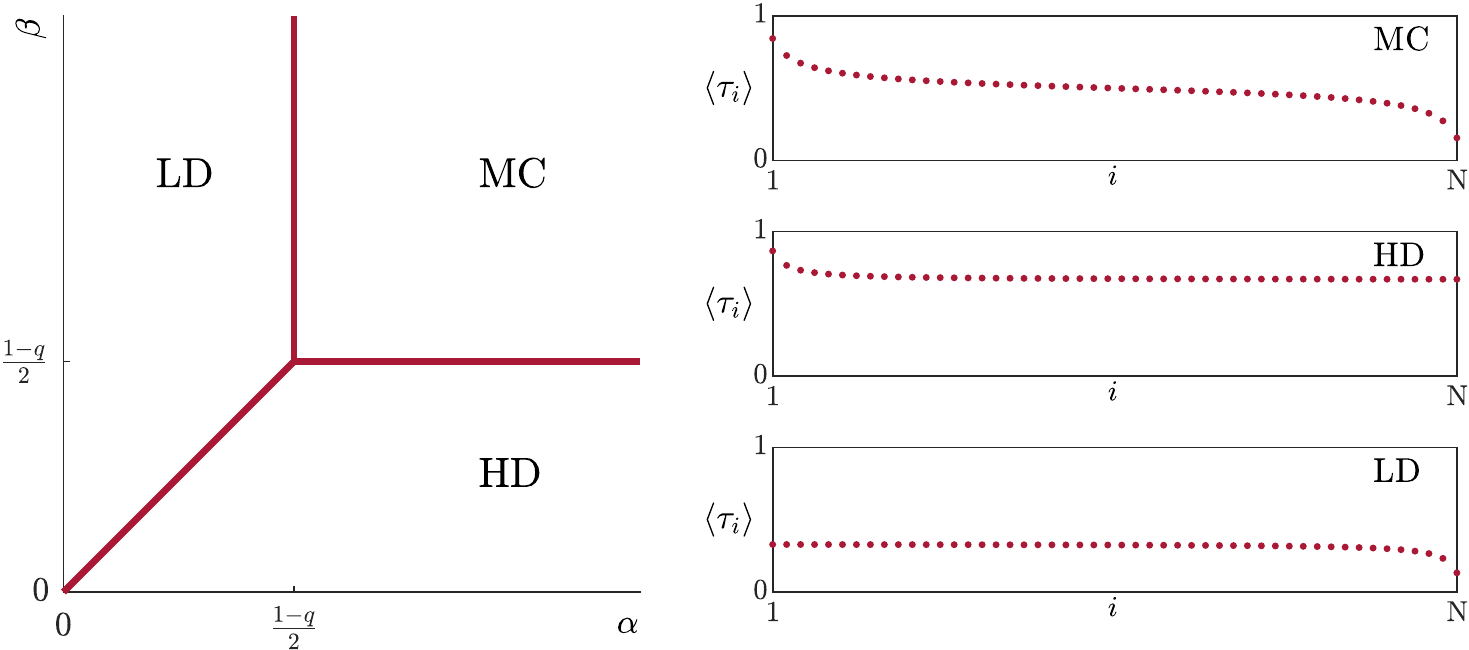}
	\caption{Phase diagram of the TASEP and typical density profiles $\langle\tau_i\rangle$ in each of the three phases.}
	\label{fig:PhaseDiagram}
\end{figure}

For $0\leq q < 1$, the ASEP exhibits three phases that depend on $\alpha$, $\beta$, $q$. The transitions between phases are typified by sharp changes (nonanalyticities in the limit $N\to\infty$) in the steady-state particle current $J$ and density profile $\langle \tau_i\rangle$. See Figure~\ref{fig:PhaseDiagram} for the phase diagram and plots of the density profile in the three phases.

In detail, the phases are:
\begin{itemize}
	\item \emph{High density} (HD), $\alpha > \beta$, $\beta < (1-q)/2$. Here, $J \sim \beta(1-q-\beta)/(1-q)$. The slow rate of exit from the lattice, $\beta$, creates an accumulation of particles, queueing to leave the system at the right hand boundary, which propagates back into the bulk of the system.
Consequently the system is characterised by a high occupation density overall.

	\item \emph{Low density} (LD), $\alpha < \beta$, $\alpha < (1-q)/2$. Here, $J \sim \alpha(1-q-\alpha)/(1-q)$. The particles enter the system at a low rate $\alpha$ and are able to freely move through the bulk and exit without jamming. Consequently the system is characterised by a low occupation density overall.
	\item \emph{Maximal current} (MC), $\alpha$, $\beta > (1-q)/2$. Here, the current $J \sim (1-q)/4$ is maximised. In this phase, particles enter and leave the system at high rates, and the current is restricted by how efficiently the particles can move through the bulk.
\end{itemize}

We refer the reader to \cite{blythe2007nonequilibrium} for further discussion of the phase diagram. Notice that in the limit $q \to 1$ the phase boundaries approach the axes. Consequently, the SSEP exhibits no phase transitions. We also mention the \emph{reverse bias} case $q>1$. The reduction relations 
(\ref{eq:ReductionRelations1}--\ref{eq:ReductionRelations3})
still hold in this phase (and as such the $q$-general mappings to be presented will still hold), however like the SSEP there are no phase transitions \cite{blythe2000exact}.

Finally, we have a \emph{particle-hole symmetry}: the dynamics of particles moving to the right are identical to the dynamics of holes moving to the left \cite{blythe2007nonequilibrium}. Thus the system of particles moving through the lattice from left to right with entry rate $\alpha$ and exit rate $\beta$ is identical to a system of holes moving from right to left with entry rate $\beta$ and exit rate $\alpha$. Then the high density phase is the counterpart, for holes, of the low density phase and the low density phase is the counterpart, for holes, of the high density phase.

\subsection{Explicit matrix representation}
\label{sec:explicit}

In the previous section, we saw how to use use the reduction relations (\ref{eq:ReductionRelations1}--\ref{eq:ReductionRelations4}) in a formal way to calculate configurational weights---that is, without reference to any matrix explicit representation. One can go on to calculate physical observables, such as the current and density profile, in this way \cite{derrida1993exact}. However, for the purposes of identifying mappings to combinatorial enumeration problems, it is often helpful to write out $D$, $E$, $\bra{W}$, $\ket{V}$ explicitly. Generally, no finite-size matrices obey (\ref{eq:ReductionRelations1}--\ref{eq:ReductionRelations4}) (except along special parameter curves \cite{Essler_1996,Mallick_1997}) and we instead resort to semi-infinite representations, of which several are known. For our purposes, the most useful representation is \cite{sasamoto1999orthogonal,blythe2000exact,blythe2007nonequilibrium}:
\begin{eqnarray}
\label{eq:ExplicitMatrixRep1}
\fl D &= \frac{1}{1-q}\left( \begin{array}{ccccc}
1 & \sqrt{1-q} & \cdot & \cdot & \cdots \\
\cdot & 1 & \sqrt{1-q^2} & \cdot & \cdots \\
\cdot & \cdot & 1 & \sqrt{1-q^3} & \cdots \\
\cdot & \cdot & \cdot & 1 & \cdots \\
\vdots & \vdots & \vdots & \vdots & \ddots \end{array} \right) \\
\fl E &= \frac{1}{1-q}\left( \begin{array}{ccccc}
1 & \cdot & \cdot & \cdot & \cdots \\
\sqrt{1-q} & 1 & \cdot & \cdot & \cdots \\
\cdot & \sqrt{1-q^2} & 1 & \cdot & \cdots \\
\cdot & \cdot & \sqrt{1-q^3} & 1 & \cdots \\
\vdots & \vdots & \vdots & \vdots & \ddots \end{array} \right)  \label{eq:ExplicitMatrixRep2}
\end{eqnarray}
with the boundary vectors
\begin{eqnarray}
\fl \bra{W} &= \kappa\left(1, \frac{a}{\sqrt{1-q}},\;\frac{a^2}{\sqrt{(1-q)(1-q^2)}},\;\frac{a^3}{\sqrt{(1-q)(1-q^2)(1-q^3)}}, \dots \right) \label{eq:ExplicitMatrixRep3} \\
\fl \ket{V} &= \kappa\left(1, \frac{b}{\sqrt{1-q}},\;\frac{b^2}{\sqrt{(1-q)(1-q^2)}},\;\frac{b^3}{\sqrt{(1-q)(1-q^2)(1-q^3)}}, \dots \right)^T \label{eq:ExplicitMatrixRep4} \;.
\end{eqnarray}
$\kappa$ is a normalisation factor defined such that $\braket{W}{V} = 1$. Here we have employed the shorthand
\begin{equation}
\label{eq:abdefinition}
a = \frac{1-q-\alpha}{\alpha}\;,\qquad b = \frac{1-q-\beta}{\beta} \;.
\end{equation}
Although the objects in (\ref{eq:ExplicitMatrixRep1}--\ref{eq:ExplicitMatrixRep4})  diverge as $q\to1$,  any product of the form $\lim_{q\to1}\bra{W}DEDDE\dots\ket{V}$ is well defined, and the limit gives the correct SSEP configuration weight. Other representations are possible, see \cite{blythe2007nonequilibrium}. The matrices \eqref{eq:ExplicitMatrixRep1}, \eqref{eq:ExplicitMatrixRep2} are reminiscent of \emph{ladder operators} for the quantum harmonic oscillator. That is, these matrices act on a state ket 
\begin{equation}
\ket{n} = (\underbrace{0,0,0}_n,1,0,0, \dots)^T
\end{equation} 
to transform it into a superposition of $\ket{n}$ and $\ket{n \pm 1}$. In fact, these matrices can be related to the ladder operators for a $q$-deformed quantum harmonic oscillator, a fact that was exploited in \cite{sasamoto1999orthogonal,blythe2000exact} to calculate physical properties for the PASEP.

We now make three further observations of the ASEP, all of which motivate a combinatorial interpretation of the matrix product representation.

\subsection{Three observations}
\label{sec:ThreeObservations}

\subsubsection{Combinatorial factors in $\alpha = \beta = 1$ partition functions.} \label{sec:PartitionFunctionAlphaBeta1}

Beginning with the SSEP, we first directly calculate the partition function $Z_N$. Several methods are known \cite{derrida2002large, sasamoto1996one}, and here we present perhaps the most straightforward of these, following \cite{vanicat2017exact}. The partition function is written
\begin{equation}
\label{eq:SSEPNormalisationIntermediate}
Z_N = \bra{W}C^N\ket{V} = \bra{W}(D+E)C^{N-1}\ket{V} \;,
\end{equation}
where $C = D+E$. Now, the commutation property \eqref{eq:ReductionRelations1} for $q=1$,
\begin{equation}
[D,C] = D(D+E)-(E+D)D = C\;,
\end{equation}
implies that $DC = C(D+1)$. Repeating this $(N-1)$ times gives
\begin{equation}
DC^{N-1} = C^{N-1}(D+N-1)\;,
\end{equation}
which we insert into \eqref{eq:SSEPNormalisationIntermediate} and apply the reduction relations
\begin{eqnarray}
Z_N &= \bra{W}C^{N-1}(D+N-1)\ket{V} + \bra{W}EC^{N-1}\ket{V} \\
& = \left(\frac{1}{\alpha}+\frac{1}{\beta}+N-1\right)Z_{N-1} \\
& = \frac{\Gamma(\frac{1}{\alpha}+\frac{1}{\beta} + N)}{\Gamma(\frac{1}{\alpha}+\frac{1}{\beta} + N-1)} Z_{N-1} \;.
\end{eqnarray}
Given that $Z_0 = 1$, this recursion is easily solved to give
\begin{equation}
\label{eq:SSEPAlphaBetaPartitionFunction}
Z_N = \frac{\Gamma\left(\frac{1}{\alpha}+\frac{1}{\beta}+N\right)}{\Gamma\left(\frac{1}{\alpha}+\frac{1}{\beta}\right)} \;.
\end{equation}
In the case $\alpha = \beta = 1$, \eqref{eq:SSEPAlphaBetaPartitionFunction} reduces to 
\begin{equation}
\label{eq:SSEPPartitionFunctionab1}
Z_N = (N+1)! \;. 
\end{equation}

The factorial is the most familiar combinatorial number, counting the number of permutations of $(N+1)$ integers in this case. This suggests that the SSEP with $\alpha=\beta=1$ can be related to a uniform distribution over the space of permutations. In Section~\ref{sec:AlphaBeta1SSEP} we will see that this is the case.

We now turn to the TASEP (the case $q=0$). Although the reduction relation \eqref{eq:ReductionRelations1} simplifies when setting $q=0$, evaluation of the partition function proves more challenging in the case of general $\alpha$ and $\beta$. Nonetheless, it can be calculated directly using the reduction relations \cite{derrida1993exact} or by formal expansion of its generating function \cite{depken2003models, blythe2004grand, blythe2007nonequilibrium, wood2017renyi} (see Section~\ref{sec:WeightEnumeration} for an example). One finds
\begin{equation}
\label{eq:TASEPPartitionFunctionExpression}
Z_N = \sum_{p=1}^N\frac{p(2N-p-1)!}{N!(N-p)!}\sum_{q=0}^p \left(\frac{1}{\alpha}\right)^q\left(\frac{1}{\beta}\right)^{p-q} \;.
\end{equation}
This time, setting $\alpha = \beta = 1$ reduces \eqref{eq:TASEPPartitionFunctionExpression} to the summation
\begin{eqnarray}
\label{eq:TASEPPartitionFunctionExpressionab1}
Z_N &=& \sum_{p=1}^N\frac{p(p+1)(2N-p-1)!}{N!(N-p)!}\nonumber \\
 &=& \frac{2}{N} \sum_{p=1}^N\binom{p+1}{2}\binom{2N-p-1}{N-1}\nonumber \\
 &=& \frac{2}{N} \binom{2N+1}{N+2}
\end{eqnarray}
where we have used the Chu-Vandermonde identity \cite{gould1956some}
\begin{equation}
\sum_{p=-\infty}^{\infty}\binom{p+a}{c}\binom{b-p}{d} = \binom{a+b+1}{c+d+1}.
\end{equation}
Finally we obtain
\begin{equation}
Z_N = \frac{(2N+2)!}{(N+2)!(N+1)!} = C_{N+1} \label{eq:TASEPPartitionFunctionExpressionab1Catalan}
\end{equation}
where $C_{N+1}$ is the $(N+1)^\mathrm{th}$ \emph{Catalan number}. These numbers are very well-known in combinatorics, solving at least 60 counting problems \cite{stanley1999enumerative}. For example, $C_n$ is the number of ways to match $n$ pairs of brackets, accounting for all the different ways they may be nested. In \ref{sec:AppendixCatalanDerivation} we see that one can obtain \eqref{eq:TASEPPartitionFunctionExpressionab1Catalan} rather directly from the form of matrices and vectors.

To summarise, the results \eqref{eq:SSEPPartitionFunctionab1}, \eqref{eq:TASEPPartitionFunctionExpressionab1Catalan} give two integer sequences $(N+1)! = 1, 2, 6, 24, 120, 720, \dots$ and $C_{N+1} = 1, 2, 5, 14, 42, 132, \dots\,$, ubiquitous in enumerative combinatorics and now arising in this nonequilibrium physics problem.

\subsubsection{TASEP partition function in terms of bicoloured Motzkin and Dyck paths.} 
\label{sec:DyckMotzkin}

One way to show a connection between the TASEP and enumeration problems is to examine the explicit matrix and vector representations (\ref{eq:ExplicitMatrixRep1}--\ref{eq:ExplicitMatrixRep4}). Here, we use this approach to show a link to counting \emph{bicoloured Motzkin paths} and \emph{Dyck paths} \cite{brak2004nonequilibrium,blythe2004grand,blythe2004dyck}.

First, we denote a ladder operator as $g$, and state ket vectors $\ket{n}$ such that $g\ket{n} = \ket{n-1}$, $g^\dagger\ket{n} = \ket{n+1}$, with boundary condition $\ket{-1} = 0$. We also define the scalar product with a bra vector  $\braket{n}{k} = \delta_{nk}$. From this, and on setting $\alpha=\beta=1$, the representations (\ref{eq:ExplicitMatrixRep1}--\ref{eq:ExplicitMatrixRep4}) simplify to
\begin{eqnarray}
\fl D = 1+g = \left( \begin{array}{ccccc} 
1 & 1 & 0 & 0 & \cdots \\
0 & 1 & 1 & 0 & \cdots \\
0 & 0 & 1 & 1 & \cdots \\
0& 0& 0& 1 & \cdots \\
\vdots & \vdots & \vdots & \vdots & \ddots \end{array} \right) \qquad E = 1+g^\dagger= \left( \begin{array}{ccccc}
1 & 0   & 0& 0& \cdots \\
1& 1 & 0 & 0  & \cdots \\
0 & 1 & 1 & 0 & \cdots \\
0 & 0 & 1 & 1 & \cdots \\
\vdots & \vdots & \vdots & \vdots & \ddots \end{array} \right) \label{eq:AlphaBeta1LadderRep1}
 \\
\fl \bra{W} = \bra{0} = (1,0,0,\dots) \qquad \ket{V} = \ket{0} = (1,0,0,\dots)^T \label{eq:AlphaBeta1LadderRep2}
\end{eqnarray}
and the partition function
\begin{eqnarray}
Z_N &= \bra{0}(2+g+g^\dagger)^N\ket{0}
\label{Z11}
\end{eqnarray}
is an enumeration of all walks in the nonnegative plane that start and return to the origin, with $N$ steps from the set $\{ \nearrow, \searrow, \times, \cdot \}$, where `$\times$' and `$\cdot$' are distinct \emph{non-movement} steps. These are bicoloured Motzkin paths.

\begin{figure}[!t]
	\centering
	\includegraphics{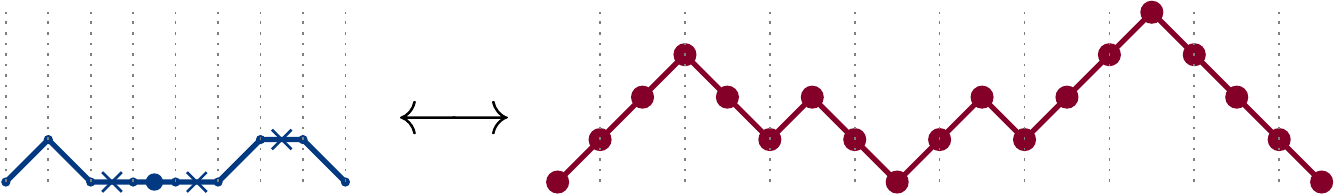}
	\caption{A bicoloured Motzkin path (left), and its equivalent Dyck path (right).
		\label{fig:DyckPath}}
\end{figure}
Each bicoloured Motzkin path of length $N$ is equivalent to a \emph{Dyck path} of length $2(N+1)$ and vice versa. Dyck paths comprise only up and down unit steps, starting and ending at the origin without going below zero. To transform a Motzkin path to a Dyck path we associate to:
\begin{itemize}
	\item each $\times$ an up step followed by a down step $(\nearrow, \searrow)$,
	\item each $\cdot$ a down step followed by an up step $(\searrow, \nearrow)$,
	\item each $\searrow$ two down steps $(\searrow, \searrow)$,
	\item each $\nearrow$ two up steps $(\nearrow, \nearrow)$,
\end{itemize}
and finally bookend each walk an with up and down step. See Figure \ref{fig:DyckPath} for an example.

\begin{figure}[!t]
	\centering
	\includegraphics{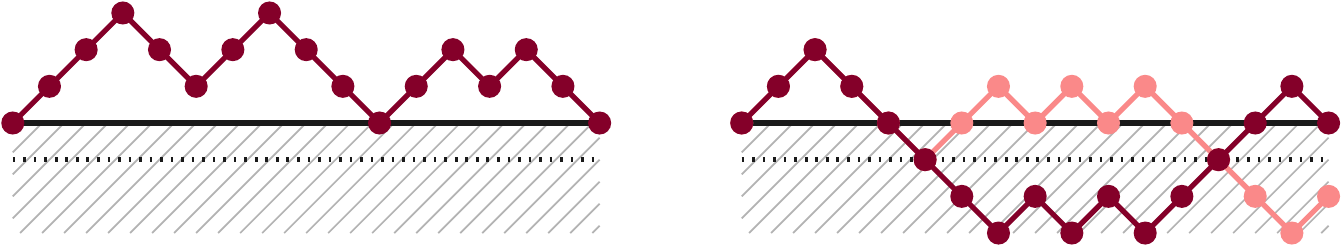}
	\caption{Left: a Dyck path, consisting of equal numbers of up-steps and down-steps such that the path never goes below $0$. Right: a walk that starts and ends at $0$, but goes below, and its reflection about the point it first touches $-1$, which then terminates at $-2$.}
	\label{fig:DyckPathReflection}
\end{figure}

The number of Dyck paths of length $2n$ is known to be the Catalan number $C_n$ \cite{stanley1999enumerative,deutsch1999dyck}. To see this note that the total number of paths that start and terminate at zero is $\left(2n\atop n\right)$. The number of \emph{invalid} paths---paths that cross below zero---is counted by reflecting these paths about the axis at the point they first hit $-1$. These reflected paths all terminate at $-2$ (Figure \ref{fig:DyckPathReflection}), and the total number of such paths is $\left(2n\atop n-1\right)$. The number of \emph{valid} paths is then $\left(2n\atop n\right) - \left(2n\atop n-1\right) = \frac{1}{n+1}\left(2n\atop n\right) = C_n$.

For the case $\alpha=\beta=1$, each of these paths are equally weighted. Therefore the normalisation, $Z_N$ just counts the total number of paths and is equal to $C_{N+1}$, consistent with Eq.~\eqref{eq:TASEPPartitionFunctionExpressionab1Catalan}. See \ref{sec:AppendixCatalanDerivation} or \cite{derrida1993exactdiff} for more details.

\subsubsection{The one-transit walk.}
\label{sec:OneTransitWalk}

\begin{figure}[!t]
	\centering
	\includegraphics{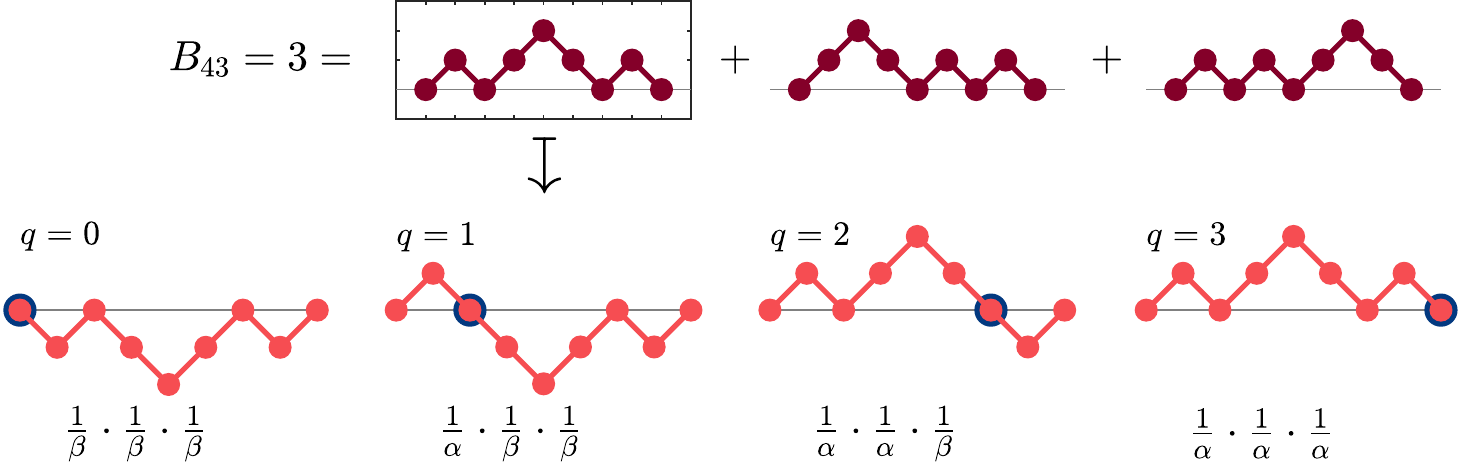}
	\caption{Dyck paths and one-transit walks. The first row illustrates the three Dyck paths with $2N = 8$ steps and $p = 3$ returns. The second row illustrates the four one-transit walks corresponding to the first Dyck path: the Dyck path inverted at each return to make a set of $p+1 = 4$ one-transit walks, each with associated weights.}
	\label{fig:OneTransitWalks}
\end{figure}

Finally, staying with the TASEP, the partition function \eqref{eq:TASEPPartitionFunctionExpression} contains the combinatorial factor
\begin{equation}
B_{Np} = \frac{p(2N-p-1)!}{N!(N-p)!}
\end{equation}
which is sometimes referred to as a \emph{ballot number} \cite{comtet2012advanced,Brak_2001,brak2004nonequilibrium}, and is the solution to the following enumerative problem: the number of Dyck paths that can be drawn of length $2N$ that return to the origin $p$ times (including the final return). An example of this is shown in Figure \ref{fig:OneTransitWalks}.

Now, for each of these walks with $p$ returns, we create a set of $(p+1)$ walks whereby the walk is \emph{inverted} about zero at the $q^{\rm th}$ return, taking $q = (0, 1, \dots p)$, again see Figure \ref{fig:OneTransitWalks}.  Finally we associate to each of these new inverted walks a factor of $(1/\alpha)^q(1/\beta)^{p-q}$. By this construction, these walks can return to the origin multiple times, but cross it {\em at most} once. Such walks have been considered in the context of the TASEP in \cite{brak2004nonequilibrium} and is called a \emph{one-transit walk}. A weight $1/\alpha$ is applied to each return from above, and $1/\beta$ to each return from below. Summing the weights over all such walks then gives the TASEP partition function in \eqref{eq:TASEPPartitionFunctionExpression} \cite{brak2004nonequilibrium,blythe2004grand}.

In this picture, we see very clearly the connection to an equilibrium partition function over an extended configuration space. Recall that in the TASEP, there are $2^N$ configurations of particles and holes. The corresponding set of one-transit walks contains $C_{N+1}$ configurations, which exceeds $2^N$: asymptotically, $C_n \sim \frac{4^n}{\sqrt{\pi} n^{3/2}}$. Each walk has a weight that can be interpreted as a Boltzmann factor; rewriting $\tilde{\alpha} = \ln \alpha$, $\tilde{\beta} = \ln \beta$, the weight for a walk with given $p$ and $q$ can be written as $\e^{-q\tilde{\alpha}-(p-q)\tilde{\beta}}$. Summing over multiple such Boltzmann-like weights gives the TASEP partition function \eqref{eq:TASEPPartitionFunctionExpression}.

As we further discuss in Section~\ref{sec:AlphaBeta1TASEP}, the mapping from TASEP configurations to one-transit walks and other combinatorial objects is one-to-many. That is, while each walk can be uniquely identified with a TASEP configuration, the converse is not true. Another way to look at this is as the TASEP defining a partitioning of an extended configuration space. The partition function is invariant under this partitioning; however other measures, such as entropies, are sensitive to it (see Section~\ref{sec:LambdaWeightEnumeration}).

\subsection{Structure of this review}
\label{sec:Remainder}
\begin{figure}[!t]
	\centering
	\includegraphics[scale=0.8]{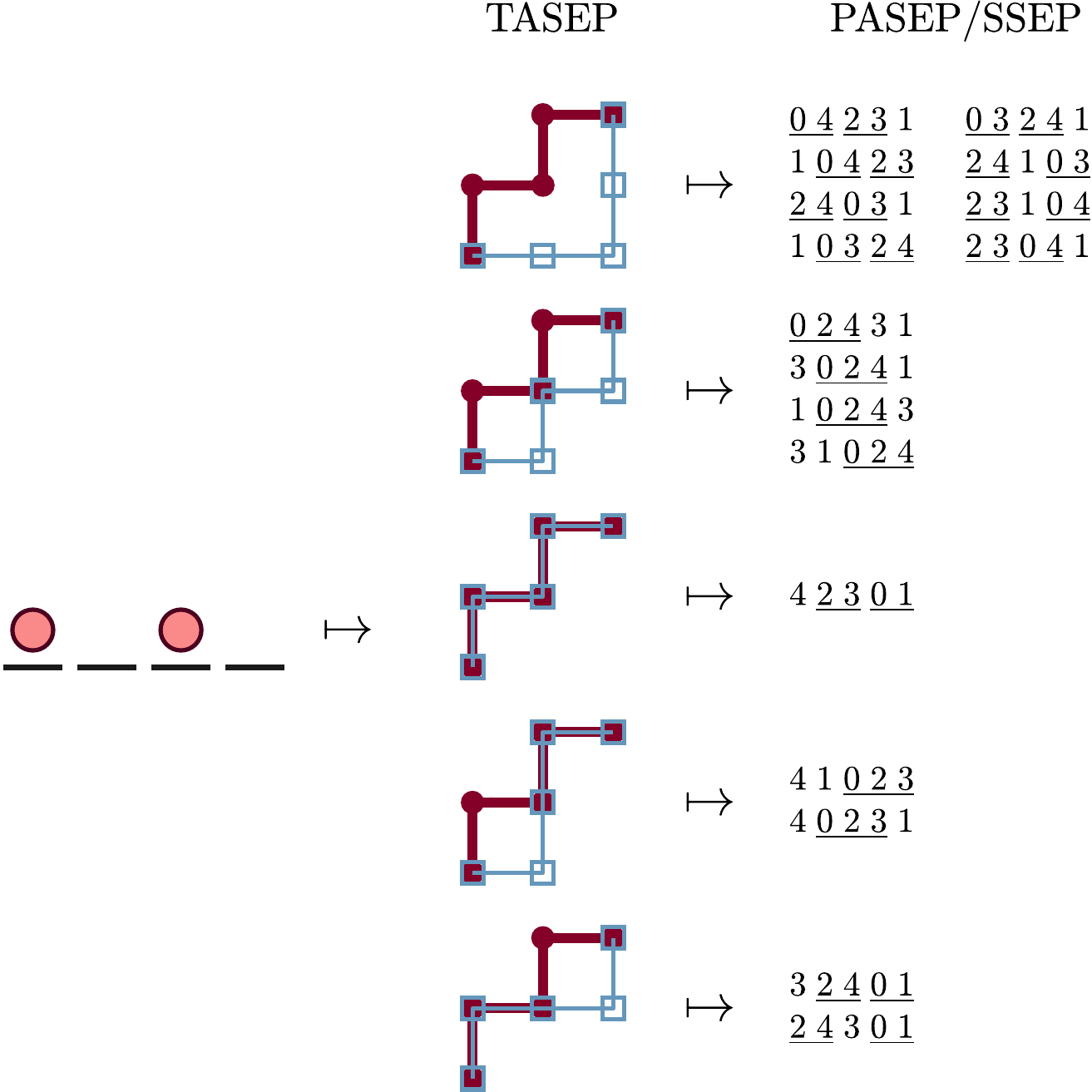}
	\caption{Schematic of the combinatorial mappings to be outlined in this review. An ASEP configuration (an example is illustrated in the left column) has a one-to-many mapping to certain \emph{dominated paths} (illustrated in the middle column), which we propose in Section \ref{sec:WeightedPermsMapping} to in turn map one-to-many to \emph{permutations} of integers that follow certain rules (illustrated in the right column).}
	\label{fig:DraftOverviewDiagram}
\end{figure}

The three observations in the previous subsection point to a deeper underlying combinatorial structure of the matrix product solution to the ASEP stationary state. In what follows we formalise and develop equivalent combinatorial interpretations of the matrix product weights. See Figure \ref{fig:DraftOverviewDiagram} for a schematic illustration of the mappings between these interpretations.

As implied in Section~\ref{sec:PartitionFunctionAlphaBeta1}, we see the combinatorial structure is at its clearest in the  $\alpha = \beta = 1$ TASEP where Catalan numbers arise. The case $\alpha = \beta = 1$, $q=0$  is the focus of Section~\ref{sec:AlphaBeta1TASEP}, where we show a mapping between nonequilibrium configurations and path enumeration problems. In Section~\ref{sec:AlphaBeta1SSEP} we discuss a combinatorial problem of \emph{permutations} that the SSEP ($q=1$) maps to. 
In Section~\ref{sec:GeneralAlphaBetaqDiscussion} we show how these mappings generalise to the full $\alpha$, $\beta$, $q$ parameter space. It turns out one can associate a $q$-dependent weight to the permutations of Section~\ref{sec:AlphaBeta1SSEP}, thus generalising to the PASEP.  It then remains to encode the other two parameters $\alpha$, $\beta$ into these mappings, which we discuss in Section~\ref{sec:AlphaBetaGeneral}.

In Section~\ref{sec:LambdaWeightEnumeration} we consider a more general calculation in the TASEP, of a quantity $H_\lambda$ known as the \emph{R\'enyi entropy}. This is a measure of the partitioning of the state space, and $\e^{H_\lambda}$ is in turn a measure of an \emph{effective number} of participating configurations in the NESS. For the case $\lambda = 2$ we show how the equivalence to path enumeration problems maps the calculation onto another combinatorial problem, involving enumerating random walks in the upper \emph{quadrant}. We may then  use results and techniques from the random walk literature to compute the R\'enyi entropy in this case.

Finally, in Section~\ref{sec:multi} we summarise generalisations of the matrix product solution to the case of multispecies ASEPs such as those involving second-class particles. Our aim is to point out further interesting combinatorial connections such as priority queues and connections with integrable systems and algebraic combinatorics that are currently being explored. We conclude in Section~\ref{sec:conc}.

\section{$\alpha = \beta = 1$ TASEP}
\label{sec:AlphaBeta1TASEP}

The $\alpha = \beta = 1$ TASEP proves to be the most analytically tractable system as the weights of configurations are integers. We outline a one-to-one mapping between configurations of the TASEP and a class of length-$N$ paths, and introduce a measure of \emph{dominance} \cite{kreweras1981solution,kreweras1965classe,niederhausen1981enumeration} to find the weight of the configuration. This mapping proves equivalent to several others, including that to Motzkin paths, which arises naturally from the explicit matrix representation (\ref{eq:ExplicitMatrixRep1}--\ref{eq:ExplicitMatrixRep4}) in the discussion of Section~\ref{sec:DyckMotzkin}. We frame the state space in terms of the path dominance mapping, as the translation from TASEP configuration to path is simple in this case and offers an intuitive link between the weight of a configuration and the area its path encloses.

\subsection{Mapping to a path dominance problem}
\label{sec:PathDominanceMapping}

Consider the set of discrete paths $\mathcal{T} \in \{\uparrow,\rightarrow\}^N$ that begin at $(0,0)$, and end at $(Q,P)$, with $P+Q=N$ steps in total. The total number of paths is
\begin{equation}
\sum_{P=0}^N \binom{N}{P} = 2^N\;. 
\end{equation}

\begin{figure}[!t]
	\includegraphics{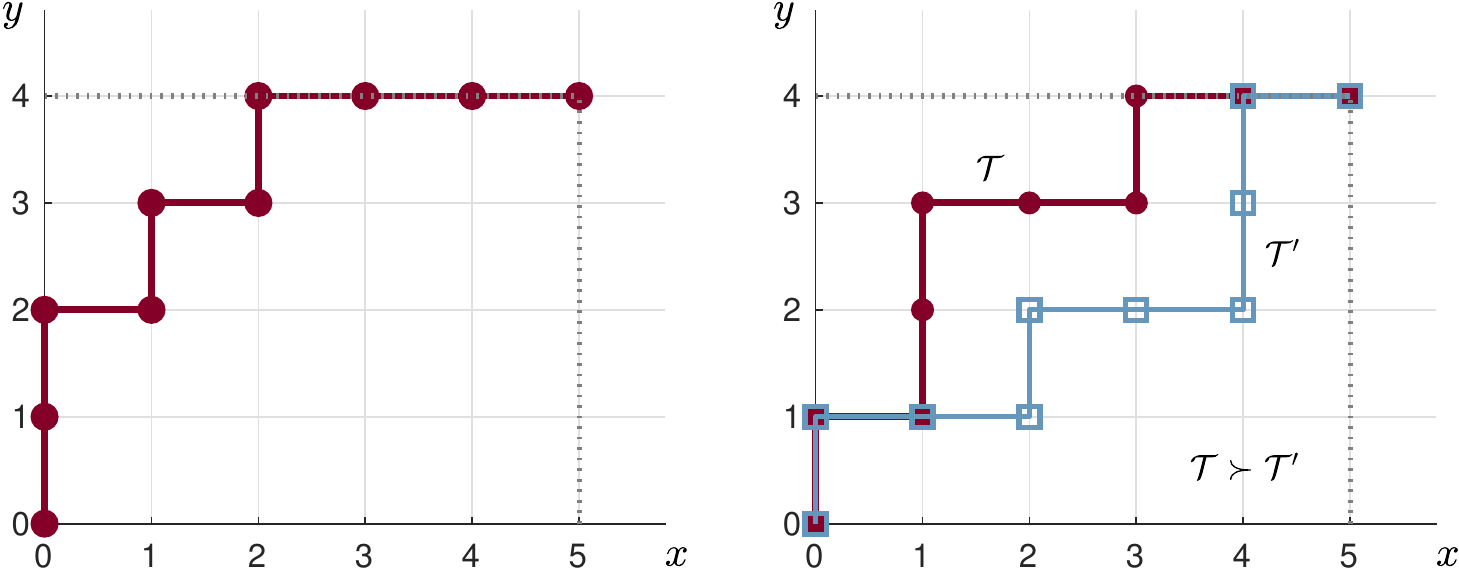}
	\caption{Left: the path $\mathcal{T}$ with the three equivalent specifications \eqref{eq:PathT1}, \eqref{eq:PathT2} and \eqref{eq:PathT3}. Right: two paths $\mathcal{T}, \mathcal{T}^\prime$. Here, $\mathcal{T}$ dominates $\mathcal{T}^\prime$.}
	\label{fig:PathDominanceExample}
\end{figure}

A path $\mathcal{T}$ can be defined by its set of steps. For example, the path shown in Figure~\ref{fig:PathDominanceExample}, left, can be specified as
\begin{equation}
\label{eq:PathT1}
\mathcal{T} = \left(\uparrow,\uparrow,\rightarrow,\uparrow, \rightarrow,\uparrow,\rightarrow,\rightarrow,\rightarrow\right) \;.
\end{equation}
Alternatively, we can specify, for each value of the $x$-coordinate $0,1,\ldots Q$ the maximal $y$-coordinate of the path. For the path $\mathcal{T}$ above, we would have
\begin{equation}
\label{eq:PathT2}
y^{(\mathcal{T})} = \left(2,3,4,4,4,4\right)\;. 
\end{equation}
Equally, we could specify the maximal $x$-coordinate for each value of the $y$-coordinate $0,1,\ldots P$:
\begin{equation}
\label{eq:PathT3}
x^{(\mathcal{T})} = \left(0,0,1,2,5\right) \; .
\end{equation}

With this formalism established, we can now define what is meant by dominance \cite{kreweras1981solution,kreweras1965classe,niederhausen1981enumeration}. Take two paths $\mathcal{T}$, $\mathcal{T}^\prime$ which both terminate at $(Q,P)$. $\mathcal{T}$ \emph{dominates} $\mathcal{T}^\prime$ (denoted $\mathcal{T} \succ \mathcal{T}^\prime$) if $\mathcal{T}^\prime$ lies completely on or below $\mathcal{T}$ (see Figure \ref{fig:PathDominanceExample}, right). In terms of the maximal $x$ and $y$ coordinates, $\mathcal{T} \succ \mathcal{T'}$ if $x_i^{(\mathcal{T})} \le x_i^{(\mathcal{T}')}$ for all $i$, or, equivalently, $y_i^{(\mathcal{T})} \ge y_i^{(\mathcal{T}')}$ for all $i$. By this definition, $\mathcal{T}$ dominates itself, and it is also possible that for two paths, neither dominates the other; if the paths cross then neither path lies completely under the perimeter of the other. We emphasize that this formalism only applies to paths of the same length that have the same start and end points.

This leads to the following combinatorial problem: how many paths $\mathcal{W}(\mathcal{T})$ in total does $\mathcal{T}$ dominate?

\begin{figure}[!t]
	\includegraphics{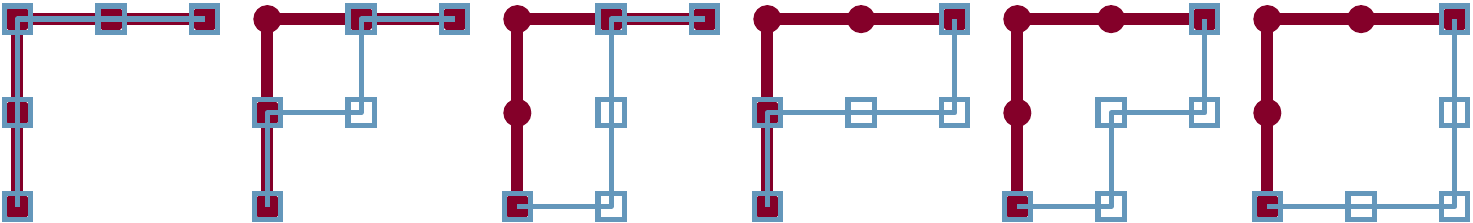}
	\caption{The weight of the path $(\uparrow, \uparrow, \rightarrow, \rightarrow)$ is $6$, as $6$ distinct paths can be drawn within its perimeter.}
	\label{fig:PathDominanceCalcExample}
\end{figure}

This quantity can be written out iteratively, accumulating all possible dominated paths as $\mathcal{T}$ grows step by step. Formally, this is
\begin{equation}
\label{eq:IterativePathWeight}
\mathcal{W}(\mathcal{T}) = \sum_{n_0 = 0}^{y_0} \sum_{n_1 = n_0}^{y_1} \sum_{n_2 = n_1}^{y_2} \cdots \sum_{n_{Q-2} = n_{Q-3}}^{y_{Q-2}} \sum_{n_{Q-1} = n_{Q-2}}^{y_{Q-1}} 1
\end{equation}
which is a set of $Q$ nested sums. Take, for example, $\mathcal{T} = (\uparrow, \uparrow, \rightarrow, \rightarrow)$, $y^{\mathcal{T}} = (2,2,2)$. This has a weight of $6$, found by manually drawing all dominated paths (Figure \ref{fig:PathDominanceCalcExample}), or from the summation
\begin{equation}
\mathcal{W}(\mathcal{T}) = \sum_{n_0=0}^2\sum_{n_1=n_0}^2 1 = \sum_{n_1 = 0}^2 1 + \sum_{n_1 = 1}^2 1 + \sum_{n_1 = 2}^2 1 = 3 + 2 + 1 = 6 \;.
\end{equation}

This problem is of interest to us as each path of length $N$ maps uniquely to a length-$N$ ASEP configuration. Specifically, an ASEP configuration with occupied sites $(j_1, j_2, \dots j_P)$ maps to a path $\mathcal{T}$ where steps $(j_1, j_2, \dots j_P)$ are $\uparrow$, and the remaining steps are $\rightarrow$. In other words, for a given dominant path $\mathcal{T}$, we can read off the TASEP configuration by going along the path and translating each upward step to a particle, and each rightward step to a hole.

The weight of a TASEP configuration then turns out to be given by $\mathcal{W}(\mathcal{T})$, the number of paths that $\mathcal{T}$ dominates. For brevity, we will also refer to $\mathcal{W}(\mathcal{T})$ as the weight of the path. For example, the path in Figure \ref{fig:PathDominanceCalcExample} maps to $\mathcal{C} = (1,1,0,0)$, which indeed has a weight of 6:
\begin{eqnarray}
\bra{W}DDEE\ket{V} &= \bra{W}D(D+E)E\ket{V} \\
& = \bra{W}DDE + DEE\ket{V} \\
& = \bra{W}\left((D(D+E))+(D+E)E\right)\ket{V} \\
& = \bra{W}(DD+D+E+D+E+EE)\ket{V} = 6 \;. 
\end{eqnarray}

\begin{figure}[!t]
	\centering
	\includegraphics{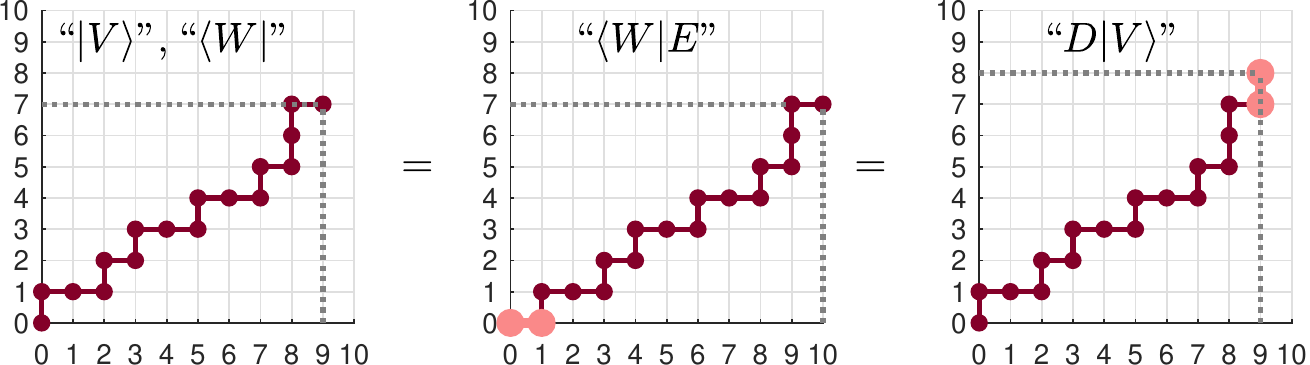}
	\caption{Graphical representation of \eqref{eq:WEPathReduction}, \eqref{eq:DVPathReduction}. Adding a $\rightarrow$ to the start or a $\uparrow$ to the end of a path does not change its weight (i.e., the number of paths it can dominate).}
	\label{fig:PathReductionRelation1}
\end{figure}

\begin{figure}[!t]
	\centering
	\includegraphics{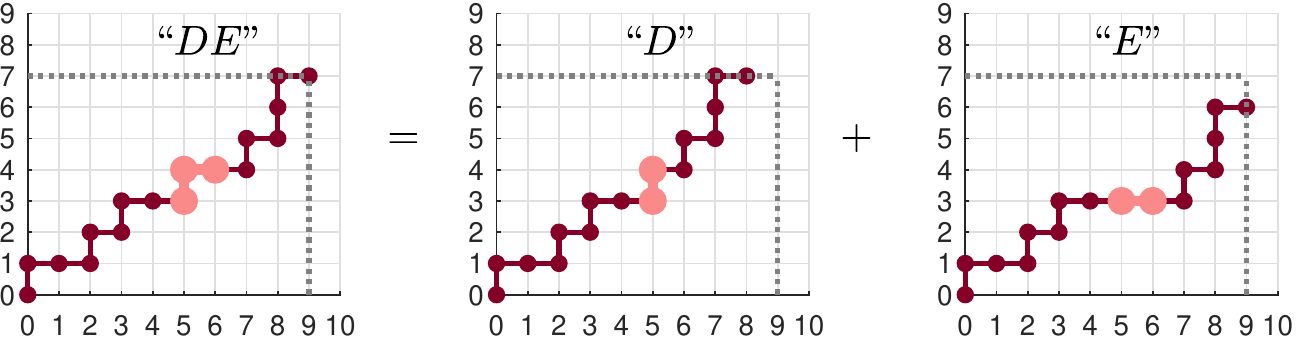}
	\caption{Graphical representation of \eqref{eq:DEPathReduction}.}
	\label{fig:PathReductionRelation2}
\end{figure}

This mapping is proven by showing that weight of a path \eqref{eq:IterativePathWeight} satisfies a set of reduction relations equivalent to (\ref{eq:ReductionRelations2}--\ref{eq:ReductionRelations1TASEP}). More formally, we require
\begin{eqnarray}
\mathcal{W}(\rightarrow,\mathcal{T}) &= \mathcal{W}(\mathcal{T}) \label{eq:WEPathReduction}\\
\mathcal{W}(\mathcal{T},\uparrow) &= \mathcal{W}(\mathcal{T}) \label{eq:DVPathReduction} \\
\mathcal{W}(\mathcal{T}_{(1)}, \uparrow,\rightarrow, \mathcal{T}_{(2)}) & = \mathcal{W}(\mathcal{T}_{(1)},\uparrow,\mathcal{T}_{(2)}) + \mathcal{W}(\mathcal{T}_{(1)},\rightarrow,\mathcal{T}_{(2)}) \label{eq:DEPathReduction}
\end{eqnarray}
where the notation $\mathcal{W}(a,b,\ldots)$ denotes concatenation of the path segments $a, b, \ldots \;$. Eqs~\eqref{eq:WEPathReduction} and \eqref{eq:DVPathReduction} are equivalent to $\bra{W}E =\bra{W}$ and $D\ket{V} = \ket{V}$ respectively, and are trivial by inspection (Figure \ref{fig:PathReductionRelation1}). Relation \eqref{eq:DEPathReduction} is the equivalent of $DE = D + E$ (see Figure \ref{fig:PathReductionRelation2}) and requires more work, but can be derived by brute force from the summation formula \eqref{eq:IterativePathWeight}, see \ref{sec:AppendixReductionDerivation}.

We now highlight three results that first originated in the path dominance literature and that we can exploit to give insights into the TASEP without any additional work. A fourth result concerns the $\lambda = 2$ R\'enyi entropy, which we reserve for Section~\ref{sec:SumSquaresDominance}. 

\subsubsection{Most probable configuration.}
The first result is a simple observation, and is that for a length-$N$ path containing $P$ $\uparrow$ steps, the most dominant path is $\mathcal{T^*} = (\uparrow, \dots \uparrow, \rightarrow, \dots \rightarrow)$, with weight
\begin{equation}
\mathcal{W}(\mathcal{T^*}) = \binom{N}{P} \;,
\end{equation}
as this rectangular path encloses all others. The equivalent TASEP configuration $\mathcal{C} = (1,\dots 1, 0,\dots 0)$ is $P$ particles stacked to the left, and is the most probable configuration with $P$ particles. Furthermore, the most probable configuration overall will be $N/2$ particles followed by $N/2$ holes (if $N$ is odd, the $\lceil N/2 \rceil$ and $\lfloor N/2 \rfloor$-particle configurations are equally most probable). In the matrix formalism, this weight corresponds to the decomposition of the string $\bra{W}D^{P}E^{N-P}\ket{V}$ using the algebraic rules.

At the other extreme, any configuration with $P$ particles stacked to the right has the minimum weight of $1$. This is because the only path that $\mathcal{T}^* = (\rightarrow, \dots \rightarrow, \uparrow, \dots \uparrow)$ dominates is itself, though in the matrix product formulation this is already trivial given $\bra{W}E\dots E D\dots D \ket{V} = 1$.

\subsubsection{Weight with fixed particle number and Narayana numbers.}
\label{sec:Narayana}
Given this mapping, the total weight of configurations $\mathcal{C}_P$ with $P$ particles is the \emph{total} weight of \emph{all} paths that terminate at $(N-P,P)$. In the path dominance literature this is known \cite{niederhausen1981enumeration}: 
\begin{eqnarray}
\label{eq:NarayanaNumber}
\sum_{\mathcal{C}_P} \mathcal{W}(\mathcal{C}_P) & = \frac{N!(N+1)!}{(N-P)!(N-P+1)!P!(P+1)!} \label{eq:FixedParticleNarayana} \\
& = \binom{N}{P}^2-\binom{N}{P+1}\binom{N}{P-1} = T(N+1,P+1)
\end{eqnarray}
where $T(n,k)$ is a \emph{Narayana number} \cite{narayana1955combinatorial} (Table \ref{tab:Narayana}, sequence A001263 in the OEIS \cite{oeisNarayana}). One combinatorial interpretation of these numbers is the number of ways to match $n$ pairs of brackets with $k$ innermost pairs. That is, the sequence $(()())$ would be counted by $T(3,2)$. Recall that the total number of ways to match $n$ pairs of brackets (without the restriction on innermost pairs) is the Catalan number $C_{n}$. Consequently $\sum_{k=1}^n T(n,k) = C_n$, and we find from \eqref{eq:NarayanaNumber} that
\begin{equation}
Z_N = \sum_{P=0}^N \left(\sum_{\mathcal{C}_P} \mathcal{W}(\mathcal{C}_P)\right) = \sum_{P=0}^N T(N+1,P+1) = C_{N+1}
\end{equation}
as previously. In \ref{sec:AppendixCatalanDerivation} we provide a derivation of these results within the matrix product formalism.

\begin{table}[]
	\centering
	\begin{tabular}{|c|ccccccc|c|c|}
		\cline{1-8} \cline{10-10}
		\backslashbox{$n$}{$k$}& 1 & 2            & 3  & 4  & 5  & 6  & 7 & & $\sum$ \\ \cline{1-8} \cline{10-10} 
		1 & 1           &             &   &   &   &   &  & & 1  \\ \cline{1-8} \cline{10-10} 
		2 & 1           & 1            &   &   &   &   &  & & 2  \\ \cline{1-8} \cline{10-10} 
		3 & 1           & 3            & 1  &   &   &   &  & & 5  \\ \cline{1-8} \cline{10-10} 
		4 & 1           & 6 & 6  & 1  &   &   &   &   & 14  \\ \cline{1-8} \cline{10-10} 
		5 & 1 & 10         & 20  & 10  & 1  &   &  & & 42 \\ \cline{1-8} \cline{10-10} 
		6 & 1           & 15           & 50 & 50 & 15  & 1  &  & & 132 \\ \cline{1-8} \cline{10-10} 
		7 & 1           & 21          & 105 & 175 & 105 & 21 & 1 & & 429 \\ \cline{1-8} \cline{10-10} 
	\end{tabular}
	\caption{The first few Narayana numbers $T(n,k)$ \eqref{eq:NarayanaNumber}. Row sums give the Catalan numbers.}
	\label{tab:Narayana}
\end{table}

\subsubsection{Determinant form of configuration weight.}

Finally, and most significantly, Narayana \cite{narayana1955combinatorial} (and later Kreweras \cite{kreweras1965classe}) has shown in this path dominance problem that the weight of a path can be written as a determinant:
\begin{equation}
\label{eq:PathDeterminantFormulaY}
\fl\mathcal{W}(\mathcal{T}) = \det\; M \;, \qquad M_{nm}=\binom{y_{m-1}+1}{1+n-m} \;,\qquad n,m = 1,2,\dots Q\;,
\end{equation}
or equivalently (`turning the path on its side')
\begin{equation}
\label{eq:PathDeterminantFormulaX}
\fl\mathcal{W}(\mathcal{T}) = \det \; M' \;, \qquad M'_{nm} = \binom{Q-x_{P-m}+1}{1+n-m}\;, \quad n,m = 1,2,\dots P \;.
\end{equation}
With the mapping from paths, this in turn provides an analytic formula for the weight of any TASEP configuration. For example, recall the example path from Eq.~\eqref{eq:PathT1}, Figure \ref{fig:PathDominanceExample}. Using the first determinant formula, this has weight from its $y$-coordinates \eqref{eq:PathT2}
\begin{equation}
\mathcal{W}(\mathcal{T}) = \det\left(
\begin{array}{ccccc}
3 & 1 & \cdot & \cdot & \cdot \\
3 & 4 & 1 & \cdot & \cdot \\
1 & 6 & 5 & 1 & \cdot \\
\cdot & 4 & 10 & 5 & 1 \\
\cdot & 1 & 10 & 10 & 5 \\
\end{array}
\right) = 117 \label{eq:DeterminantExampleY}
\end{equation}
and equivalently using its $x$-coordinates \eqref{eq:PathT3}
\begin{equation}
\mathcal{W}(\mathcal{T}) = \det\left(
\begin{array}{cccc}
4 & 1 & \cdot & \cdot \\
6 & 5 & 1 & \cdot  \\
4 & 10 & 6 & 1 \\
1 & 10 & 15 & 6 \\
\end{array}
\right) = 117 \;. \label{eq:DeterminantExampleX}
\end{equation}
This path $\mathcal{T}$ maps to the TASEP configuration
\begin{equation}
\mathcal{C} = (1,1,0,1,0,1,0,0,0)
\end{equation}
which then implies that the determinants in \eqref{eq:DeterminantExampleY} and \eqref{eq:DeterminantExampleX} are equivalent to the matrix product 
\begin{equation}
\mathcal{W}(\mathcal{C})= \bra{W}DDEDEDEEE\ket{V} \;.
\end{equation}
This reveals a deeper link between the matrix product approach and the reduction relations (\ref{eq:ReductionRelations2}--\ref{eq:ReductionRelations1TASEP}) with an elegant determinant structure.
Probing these determinants further, notice from this example that reading down each column reveals the $(y_{m-1}+1)^{\mathrm{th}}$ row in Pascal's triangle. It is also `nearly' a lower-diagonal matrix, and a simple example of a \emph{Hessenberg} matrix. Taking \eqref{eq:DeterminantExampleY}, this gives a simplified recursive determinant formula (adapted from Theorem 2.1 of \cite{kaygisiz2013determinants}):
\begin{eqnarray}
\det \; M &= \sum_{r=1}^Q (-)^{Q-r}M_{Qr}\;\det\; M_{(r-1)} \label{eq:HessenbergRecurrence} \\
&= \sum_{r=1}^Q (-)^{Q-r}\binom{y_{r-1}+1}{Q-r+1}\det\; M_{(r-1)}
\end{eqnarray}
where $M_{(r-1)}$ is the $(r-1)^{\mathrm{th}}$ minor of $M$.

In the context of the TASEP, this determinant formula has since been improved upon to encode $\alpha$ and $\beta$, see Section~\ref{sec:AlphaBetaDeterminant} and \cite{mandelshtam2015determinantal}.

\subsection{Other representations}

We refer to an ordered pair of two paths where one dominates the other as a \emph{dominated path}. As previously noted in Section~\ref{sec:Narayana}, the total number of dominated paths is given by the Catalan number $C_{N+1}$. This set of dominated paths is the extended configuration space. This space can be equivalently expressed in terms of bicoloured Motzkin paths or ``complete configurations'', which we now discuss.

\subsubsection{Bicoloured Motzkin paths.}
\label{sec:MotzkinWalks}

From the matrix representation (\ref{eq:ExplicitMatrixRep1}--\ref{eq:ExplicitMatrixRep4}), bicoloured Motzkin paths naturally arise \cite{brak2004asymmetric,blythe2004dyck}. Here we establish the link between these walks and the dominated path formalism.

The full partition function $Z_N$ is the number of unique dominated paths. Consider one such configuration with two paths $\mathcal{T} \succ \mathcal{T}'$. Denote the $i^{\rm th}$ steps of $\mathcal{T}$, $\mathcal{T}'$ as $\mathcal{T}(i)$, $\mathcal{T}'(i)$ respectively.

Comparing the two paths, on each step we have four possible outcomes, which we track with a height difference $h \geq0$, that must start and end at zero:
\begin{itemize}
	 \item $\mathcal{T}(i)=\;\uparrow$ and $\mathcal{T}'(i) =\;\rightarrow$. The paths diverge, $\Delta h = 1$,
	 \item $\mathcal{T}(i) = \;\rightarrow$ and $\mathcal{T}'(i) = \;\uparrow$. The paths converge, $\Delta h = -1$,
	 \item $\mathcal{T}(i) = \mathcal{T}'(i) =\; \uparrow$. The paths run parallel vertically, $\Delta h = 0$,
	 \item $\mathcal{T}(i) = \mathcal{T}'(i) = \; \rightarrow$. The paths run parallel horizontally, $\Delta h = 0$.	
\end{itemize}
\begin{figure}[!t]
	\includegraphics{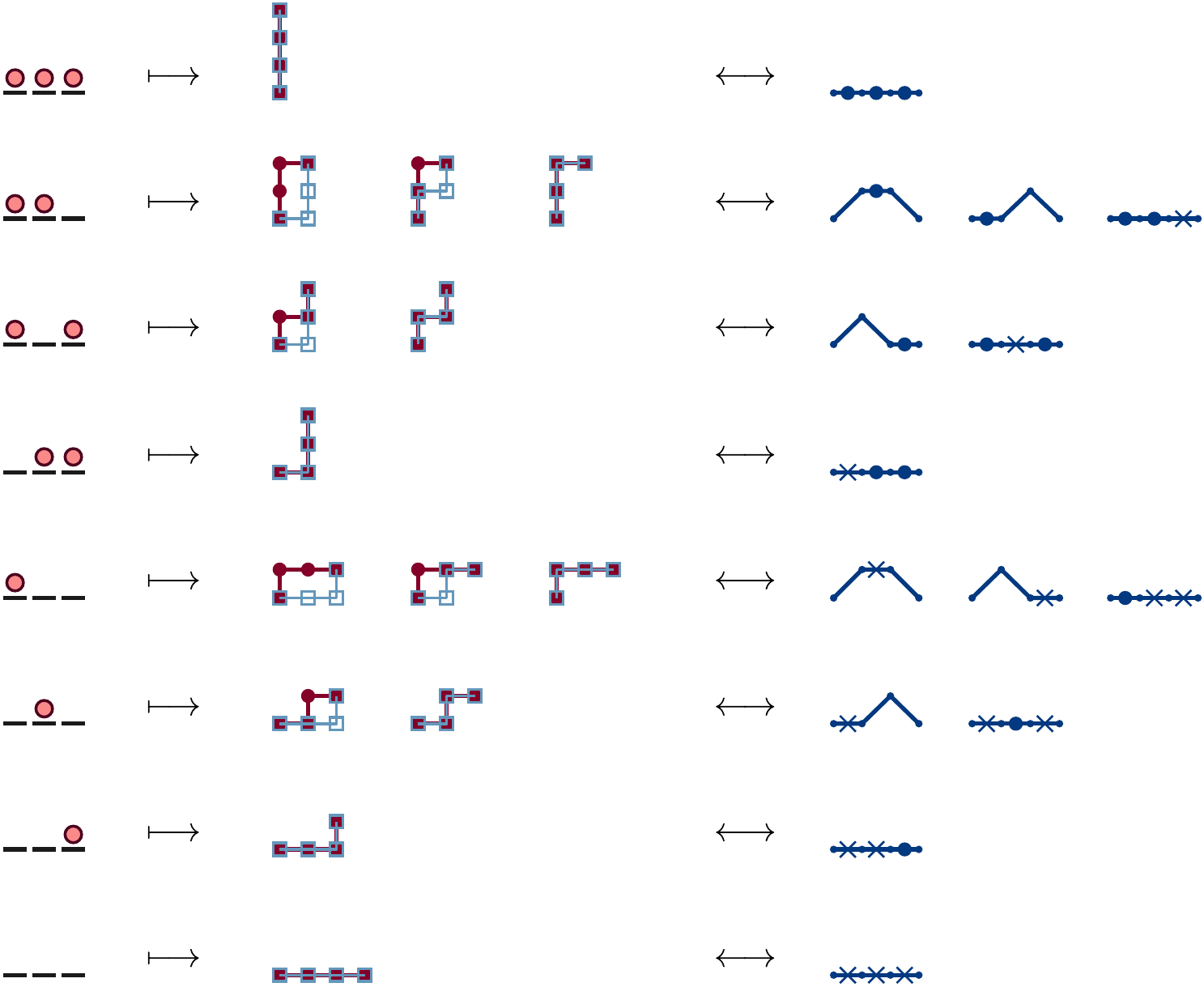}
	\caption{Calculation of the TASEP partition function for $N=3$. For each configuration (left), we draw draw all combinations of length-$N$ paths that dominate another (centre), and their equivalent bicoloured Motzkin path (right).}
	\label{fig:Z3MotzkinDominanceMap}
\end{figure}

Over each step, $h$ can therefore change by $\pm 1$, or zero in two distinct ways (denoted with `$\cdot$' and `$\times$'). The partition function is then equivalently the number of paths moving left-to-right of length $N$, from the step set $\{\nearrow, \searrow, \;\cdot\;, \times\}$, that start and end at zero, without going below zero (as $\mathcal{T} \succ \mathcal{T}'$). This is then an enumeration of bicoloured Motzkin paths. 

Extending this idea, the weight of a length-$N$ configuration $\mathcal{C}$ with sites $(j_1, j_2, \dots j_P)$ occupied is the number of length-$N$ bicoloured Motzkin paths that can be drawn from $\{\nearrow, \cdot\}$ at steps $(j_1, j_2, \dots j_P)$, and $\{\searrow, \times\}$ in the remaining steps. Thus each Motzkin path is mapped one-to-one to a dominated path. See Figure \ref{fig:Z3MotzkinDominanceMap} for an example, where we draw all $N=3$ ASEP configurations in terms of Motzkin paths.

This Motzkin path interpretation aligns neatly with the explicit representation we quote in (\ref{eq:ExplicitMatrixRep1}--\ref{eq:ExplicitMatrixRep4}). For other explicit representations, other path interpretations naturally arise. Brak et al.\ present a comprehensive set of these alternative walks in \cite{brak2004asymmetric}, as well as encoding weights to generalise for $\alpha$, $\beta$, $q$. 

\subsubsection{Markov chain of ``complete configurations''.}
\label{sec:completeconfigs}

Duchi and Schaeffer \cite{duchi2005combinatorial} express this same space of $C_{N+1}$ configurations as a set of closed, \emph{two}-row systems, which they term \emph{complete configurations}. Furthermore, they define a Markov process in this space that reproduces ASEP dynamics on the top row of the system. 

Each of these complete configurations comprise $N$ particles and $N$ holes (which they refer to as ``black'' and ``white'' particles), arranged across two rows of length $N$. The particles may be arranged in any way across both rows, given the constraint that there are always at least as many particles as holes in the first $i$ columns, $i = 1, 2, \dots N$ (the ``positivity condition''). The top-row configuration is the ASEP state that the complete configurations map to. 

The Markov process that the authors construct is a clockwise flow of these $N$ particles around both rows, with ASEP-like hopping on the top row, and a set of bottom-row dynamics (involving long-range ``sweeps'' of clusters of particles or holes) so as to preserve the positivity condition. The top row of these closed configurations replicate open TASEP dynamics. In particular, we note that a feature of a complete configuration is that if the top-left site is empty, the bottom-left row is occupied---otherwise the positivity condition would be violated. This means that a particle can always enter the top row at a rate $\alpha$, just as in the TASEP. Similarly, if the top-right site is occupied, the bottom-right site must be empty, allowing particles to exit the top row at rate $\beta$, again as in the TASEP.

\begin{figure}[!t]
	\centering
	\includegraphics{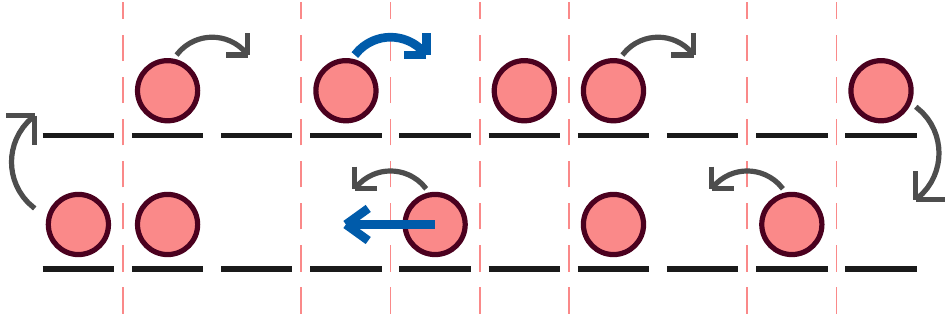}
	\caption{Simplified two-row dynamics for the case $\alpha=\beta =1$ which generates a uniform distribution over the space of complete configurations. Particles hop clockwise into empty spaces around the lattice at unit rate. Vertical dashed lines indicate zone boundaries, wherein each zone contains an equal number of particles and holes. If, on the upper row, a particle crosses a zone boundary when it hops (shown by the curved blue arrow), the particle below the receiving site is also forced to hop (shown by the straight blue arrow).}
	\label{fig:SimplifiedTwoRow}
\end{figure}

In the general $\alpha$, $\beta$ case, the bottom row dynamics that yields the desired weighting of configurations in the extended space is rather complex. However, there is a simplified dynamics, involving only local moves, that generates a uniform distribution over the extended set of complete configurations in the special case where $\alpha=\beta=1$. These dynamics, which we derive in \ref{sec:AppendixSimplifiedTwoRow}, are illustrated in Figure~\ref{fig:SimplifiedTwoRow}. Each particle hops at unit rate, as long as the site in front of them is empty. As noted above, the positivity condition already guarantees that particles enter and leave the top row at unit rate.

The one nontrivial aspect of the dynamics involves the notion of \emph{zones}. Going from left to right across the lattice, we mark a zone boundary at each point where the total number of particles to the left of the zone boundary (on both rows, and all the way back to the left boundary) is equal to the total number of holes. These are shown with vertical dashed lines in Figure~\ref{fig:SimplifiedTwoRow}. Now, if a particle on the top row can hop across a zone boundary, then the positivity condition implies that the site below must be empty, and that the site to the right on the bottom row must contain a particle. If the particle on the top row were to hop, then the positivity condition would be violated. Thus, to preserve positivity, the bottom row particle to the right must hop to the left when the particle on the top row hops. This forced bottom row move is shown with a straight arrow in Figure~\ref{fig:SimplifiedTwoRow}. Note that if we cover up the bottom row, we obtain the usual dynamics of the TASEP. However, if we cover up the top row, we get a different dynamics as a bottom row particle may move at double the usual rate, depending on the configuration of the top row.

By construction, the dynamics in the full system remain within the space of complete configurations. It can also be shown that every complete configuration can be reached from any other (see \ref{sec:AppendixSimplifiedTwoRow}). This means that all complete configurations have a nonzero probability in the steady state. Finally, the number of ways out of each configuration is equal to the number of ways in (see again \ref{sec:AppendixSimplifiedTwoRow}). Together, these three properties imply that the distribution over complete configurations is uniform. One can also show that these dynamics are dynamically reversible \cite{kelly1979reversibility}, that is, they satisfy a generalisation of detailed balance (see \ref{sec:AppendixSimplifiedTwoRow}). When we sum over the bottom row configurations for any given top row, we obtain the weight of a physical configuration of the TASEP.
\begin{figure}[!t]
	\centering
	\includegraphics{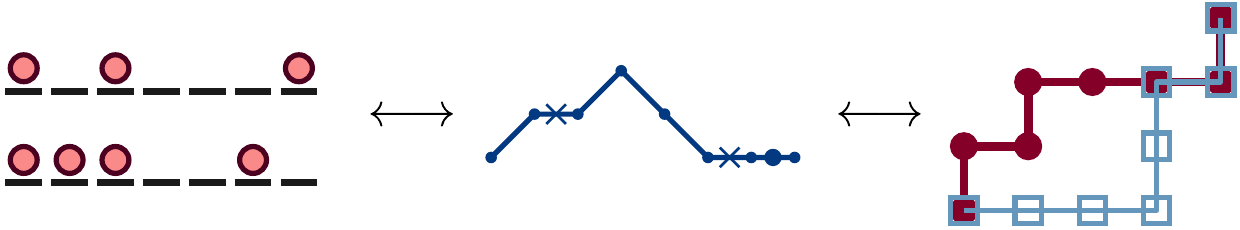}
	\caption{Example of a complete configuration in \cite{duchi2005combinatorial} (left) and its equivalent Motzkin path (centre) and dominated path (right). The top row of the complete configuration shows that these correspond to $\mathcal{C} = (1,0,1,0,0,0,1)$.}
	\label{fig:DuchiCompleteConfiguration}
\end{figure}

The basic principles that apply to this simplified dynamics can also be shown for the general $\alpha$, $\beta$ case \cite{duchi2005combinatorial}. In particular, one can uniquely associate to each way into a configuration a way to leave it. Not all of the moves take place at the same rate when $\alpha$ or $\beta$ is not equal to $1$; consequently complete configurations have different weights in this case (see Section~\ref{sec:GeneralAlphaBetaqDiscussion}).

Here we expand on how each complete configuration in this two-row system maps to a Motzkin path or dominated path. With reference to Figure \ref{fig:DuchiCompleteConfiguration}, if we assign to each column with $(\tau_{\mbox{\tiny top}};\tau_{\mbox{\tiny bottom}})$ entries a $\nearrow$ for $(1;1)$, a $\searrow$ for $(0;0)$, a `$\cdot$' for $(1;0)$ and `$\times$' for $(0;1)$, then configurations are once again a set of bicoloured Motzkin paths once the positivity condition is imposed. Corteel and Williams \cite{corteel2007markov} have since introduced a Markov chain that reproduces PASEP dynamics (where the additional parameter $q$ is introduced), using a larger set of $(N+1)!$ configurations. The broader idea of connecting stochastic transport processes to simpler ``dual'' systems has also been applied to analyse models away from the ASEP, see \cite{Carinci2013}.

\section{$\alpha = \beta = 1$ SSEP}
\label{sec:AlphaBeta1SSEP}

Our discussion so far has been limited to the TASEP ($q=0$). We now move from the totally asymmetric case to the totally \emph{symmetric} case where particles can hop either direction in the bulk at equal rates, by setting $q=1$.

We previously showed that the SSEP partition function
in the case $\alpha=\beta=1$ is $Z_N = (N+1)!$, see Eq.~\eqref{eq:SSEPPartitionFunctionab1}. This combined with the analysis of the TASEP in Section~\ref{sec:AlphaBeta1TASEP} suggests that the $2^N$ configurations of the SSEP may map to an even larger set of $(N+1)! > C_{N+1} > 2^N$ configurations. This indeed turns out to be the case; consider the integers $(0, 1, \dots N)$, of which there are $(N+1)!$ permutations. The $2^N$ configurations of the SSEP define a partitioning of these $(N+1)!$ permutations.

\subsection{Mapping to a permutation problem}
\label{sec:MappingPermutationProblem}

This mapping was first identified and formally proven by Corteel and Williams \cite{corteel2007markov} in the context of a Markov chain of permutations. Here we focus only on the mapping from SSEP to permutations using a slightly different but equivalent formalism to \cite{corteel2007markov}. This is complemented with a more detailed analysis in \ref{sec:AppendixPermutationReductionRelation}.

Consider a permutation of the $(N+1)$ integers in $(0,1,\dots N)$, denoted $(i_1, i_2, \dots i_{N+1})$. Reading this string of integers left-to-right, we say that $i_n$ has been \emph{raised} by $i_{n+1}$ if $i_{n+1}>i_n$. This time, we are interested in the following problem: how many permutations are there where only a particular set of integers $(j_1, j_2, \dots j_P)$ are raised?

This proves to be equivalent to the weight of a length-$N$ SSEP configuration with particles at sites $(j_1, j_2, \dots j_P)+1$. We illustrate this with an example. The SSEP configuration $\mathcal{C} = (0,1,0,0)$, has $N=4$ sites, and $P = 1$ particle at position $j_1 + 1= 2$. This has a weight of $7$, calculated directly with the reduction relations (\ref{eq:ReductionRelations1}--\ref{eq:ReductionRelations4}):
\begin{eqnarray}
\mathcal{W}(\mathcal{C}) &= \bra{W}EDEE\ket{V} \\
&= \bra{W}E(ED+D+E)E\ket{V} \\
& = \bra{W}(EE(ED+D+E)+E(ED+D+E)+EE)\ket{V} \\ 
& = 7 \;.
\end{eqnarray}
As anticipated, there are also $7$ permutations of $(0,1,2,3,4)$ where \emph{only} $j_1 = 1$ is raised (the underline highlights where an integer has been raised):
\begin{eqnarray}
&4\;3\;\vec{1\;2}\;0, \quad 3\;2\;\vec{1\;4}\;0, \quad 3\;\vec{1\;4}\;2\;0,\quad 2\;\vec{1\;4}\;3\;0, \quad \nonumber \\
&4\;\vec{1\;3}\;2\;0, \quad \vec{1\;4}\;3\;2\;0, \quad 4\;2\;\vec{1\;3}\;0 \;.
\end{eqnarray}

If the SSEP indeed maps to these permutations, we should expect to find an equivalent set of reduction relations like that of the SSEP \eqref{eq:DEPathReduction}, which we show in \ref{sec:AppendixPermutationReductionRelation}, in fact for the more general $DE = qED + D + E$, where weights as powers of $q$ are associated to each permutation (see Section~\ref{sec:GeneralAlphaBetaqDiscussion}).

Having established this mapping, we can quickly derive the steady-state density profile and arbitrary-order correlations between sites. We also use a result in the literature on the combinatorics of permutations, which allows us to find the probability of finding $P$ particles in the system.

\subsubsection{Steady-state density profile.}
We can now identify the average steady-state occupation of site $i$
\begin{equation}
\langle \tau_i \rangle = \frac{\bra{W}(D+E)^{i-1}D(D+E)^{N-i}\ket{V}}{\bra{W}(D+E)^N\ket{V}}
\end{equation}
as being the fraction of permutations of $(0,1,\dots N)$ where integer $(i-1)$ is raised. Note that we do not care whether any other integers are raised. One slight complication is that $(i-1)$ can only be raised if it is not at the final position within the permutation. From this interpretation we can very quickly calculate the full density profile. If $(i-1)$ is not in the final position, $(i-1)$ can be raised by any of $(i, i+1, \dots N)$ from the $N$ integers greater than $(i-1)$, giving fraction $\frac{N-(i-1)}{N}$. We then multiply by the fraction of permutations where $i$ is not in the final position which is $\frac{N}{N+1}$. We thus obtain
\begin{equation}
\langle \tau_i \rangle = \frac{N+1-i}{N} \frac{N}{N+1} = 1-\frac{i}{N+1}
\end{equation}
recovering the known linear profile \cite{spohn1983long}.

\subsubsection{Arbitrary-order correlation functions.}
\label{sec:ArbOrder}
We can extend this approach to calculate higher-order correlations between different sites without having to perform any explicit matrix calculation. First, consider the correlation
\begin{equation}
\langle \tau_{i_1}\tau_{i_2} \rangle = \frac{\bra{W}(D+E)^{i_1-1}D(D+E)^{i_2-i_1-1}D(D+E)^{N-i_2}\ket{V}}{\bra{W}(D+E)^N\ket{V}} \;,
\end{equation}
where $i_2 > i_1$. This is equivalently the fraction of permutations of $(0,1,\dots N)$ where both $(i_1-1)$ \emph{and} $(i_2-1)$ are raised. 

First, $(i_2-1)$ can be raised by any of the $(N+1-i_2)$ integers from $(i_2,\dots N)$, and the fraction of suitable permutations is then $(N+1-i_2)/(N+1)$. In this subset, $(i_1-1)$ can be raised by any of the $(N+1-i_1)$ integers from $(i_1,\dots N)$, \emph{excluding} the integer that raised $(i_2-1)$. The fraction of valid permutations here is then $(N-i_1)/N$. Combined, we then recover the result from \cite{derrida2002large,spohn1983long}
\begin{equation}
\fl \langle \tau_{i_1}\tau_{i_2} \rangle = \left(\frac{N+1-i_2}{N+1}\right)\left(\frac{N-i_1}{N}\right) = \left(1-\frac{i_2}{N+1}\right)\left(1-\frac{i_1}{N}\right) \;.
\end{equation}
By the same interpretation this can be extended to an arbitrary-order correlation between $K$ different sites $i_K, i_{K-1}, \dots i_1$, where $i_K > i_{K-1}>\dots>i_1$ \cite{derrida2004current,sasamoto1996one}:
\begin{equation}
\fl \left\langle \tau_{i_K}\tau_{i_{K-1}}\dots\tau_{i_2}\tau_{i_1} \right\rangle = \prod_{k=1}^K \left(\frac{N+1+k-K-i_k}{N+1+k-K}\right) = \prod_{k=1}^K \left(1-\frac{i_k}{N+1+k-K}\right) \;.
\end{equation}

\subsubsection{Weight with fixed particle number and Eulerian numbers.}

The sum of all weights of configurations $\mathcal{C}_P$ with $P$ particles is the number of permutations of $(0,1,\dots N)$ with a total of $P$ integers raised (again we do not care which integers in particular). We state the result from the combinatorial literature \cite{worpitzky1883studien,carlitz1959eulerian}:
\begin{equation}
\label{eq:PASEPSumPParticleWeights} 
\sum_{\mathcal{C}_P} \mathcal{W}(\mathcal{C}_P) = \euler{N+1}{P}
\end{equation}
where 
\begin{equation}
 \label{eq:EulerianNumber}
\euler{n}{k} = \sum_{j=0}^{k+1}(-)^j\binom{n+1}{j}(k+1-j)^n
\end{equation}
is known as an \emph{Eulerian number} (Table \ref{tab:Eulerian}, sequence A008292 in the OEIS \cite{oeisEulerian}), and has several neat properties reminiscent of binomial coefficients, such as the recursion \cite{petersen2015eulerian}
\begin{equation}
\euler{n+1}{k} = (n+1-k)\euler{n}{k-1} + (k+1)\euler{n}{k} \;.
\end{equation}
The generating function of \eqref{eq:EulerianNumber} is succinct \cite{petersen2015eulerian}:
\begin{eqnarray}
\fl G(t,z) = \sum_{N\geq0}\sum_{P\geq0}\euler{N+1}{P}\frac{t^N z^P}{(N+1)!} = \frac{1-\e^{t(z-1)}}{t \left(\e^{t(z-1)}-z\right)} \\
\fl = 1+\frac{1}{2!} t (z+1)+\frac{1}{3!} t^2 \left(z^2+4 z+1\right)+\frac{1}{4!} t^3 \left(z^3+11 z^2+11 z+1\right) + \dots
\end{eqnarray}
where the coefficient $\{t^Nz^P\}G(t,z)$ is the probability of finding $P$ particles in a length-$N$ SSEP. Finally, the summation over Eulerian numbers for fixed $N$ is equivalent to the summation of all $N$-site SSEP weights, and gives the factorial \cite{carlitz1954eulerian}
\begin{equation}
\sum_{P=0}^N \euler{N+1}{P} = (N+1)! \;.
\end{equation}
This is trivial in the context of Eulerian numbers, as it is simply the summation of all permutations of $(N+1)$ integers.

\begin{table}[]
	\centering
	\begin{tabular}{|c|ccccccc|c|c|}
		\cline{1-8} \cline{10-10}
		\backslashbox{$n$}{$k$}& 0 & 1 & 2  & 3  & 4  & 5  & 6 & & $\sum$ \\ \cline{1-8} \cline{10-10} 
		1 & 1 & & & & &  &  & & 1  \\ \cline{1-8} \cline{10-10} 
		2 & 1 & 1 &   &   &   &   &  & & 2  \\ \cline{1-8} \cline{10-10} 
		3 & 1 & 4 & 1  &   &   &   &  & & 6  \\ \cline{1-8} \cline{10-10} 
		4 & 1 & 11 & 11  & 1  &   &   &   &   & 24  \\ \cline{1-8} \cline{10-10} 
		5 & 1 & 26 & 66  & 26  & 1  &   &  & & 120 \\ \cline{1-8} \cline{10-10} 
		6 & 1           & 57           & 302 & 302 & 57  & 1  &  & & 720 \\ \cline{1-8} \cline{10-10} 
		7 & 1           & 120          & 1191 & 2416 & 1191 & 120 & 1 & & 5040 \\ \cline{1-8} \cline{10-10} 
	\end{tabular}
	\caption{Table of the Eulerian numbers $\langle {n\atop k} \rangle$ \eqref{eq:EulerianNumber}. Row sums yield the factorials.}
	\label{tab:Eulerian}
\end{table}

\section{Generalised parameter mappings}
\label{sec:GeneralAlphaBetaqDiscussion}

Up to now, we have focused on the parameter restriction $\alpha = \beta = 1$, $q = 0,1$. To generalise for $\alpha$, $\beta$, $q$, we do not need to expand beyond the state spaces of dominated paths and permutations already introduced, however we now associate \emph{weights}, as products of $\alpha$, $\beta$, $q$. The highlight of the following is the pleasing result that a closed-form formula is available for the weight of a general TASEP configuration in the form of a matrix determinant. Finally in Section \ref{sec:WeightedPermsMapping} we propose a mapping for general $q$ that would effectively interpolate between bicoloured Motzkin paths and weighted permutations, and in turn the weights of TASEP, PASEP and SSEP configurations.

\subsection{$\alpha = \beta = 1$ PASEP and weighted permutations}
\label{sec:qMappingPermutationProblem}

Let us recall the exact expression for the PASEP partition function \cite{blythe2000exact} which is included in \ref{sec:Zgen}. This expression interpolates between $q=1$ and $q=0$ suggesting that there may be some combinatorial entities which interpolate between the mappings we have identified in Sections \ref{sec:AlphaBeta1TASEP} and \ref{sec:AlphaBeta1SSEP}.

We first expand on the results of Section~\ref{sec:MappingPermutationProblem} by showing how an arbitrary $q$ is encoded into the mapping of SSEP configurations to permutations, first shown in \cite{corteel2007tableaux}. We remain with our slightly different formalism introduced earlier, again referring the reader to \ref{sec:AppendixPermutationReductionRelation} for further details of the mapping.

Take a permutation of the $(N+1)$ integers $(0,1,\dots N)$ in which a set of integers $(j_1, j_2, \dots j_P)$ in the permutation are raised. Let $k_\ell$ be the integer that raises $j_\ell$. For example, if the permutation contains the integer $1$ immediately followed by $4$, then we will have $j_\ell=1$ and $k_\ell=4$ for some $\ell=1,2,\dots P$. Now, for a given raise, we look for the integers with values between $j_\ell$ and $k_\ell$ and that sit to the right of the pair $j_\ell \;k_\ell$ in the permutation. Let the number of such integers be $r_\ell$. Then, to each raise we associate a weight $q^{r_{\ell}}$. The total weight of the permutation is the product of these weights.

To give a full example, consider $6\; \underline{1\;4}\;3\;\underline{0\;2\;5}$ which is one (of many) permutations that maps to the configuration $\mathcal{C} = (1,1,1,0,0,0)$. The set of raised integers $j$ is $(1,0,2)$ (corresponding to particles at sites $(2,1,3)$) and the raising integers are $(4,2,5)$. For the first pair, both the intermediate integers $2$ and $3$ lie to the right, and we acquire a weight $q^2$. For the second pair, the intermediate integer $1$ lies to the left, and we acquire the further weight $q^0=1$. For the final pair, both intermediate integers $3$ and $4$ again lie to the left, and the corresponding weight is again $q^0=1$. Combining these weights yields an overall weight of $q^2$. These weighted permutations tie into a $q$-generalisation of Eulerian numbers known as Eulerian \emph{polynomials}, introduced in \cite{williams2005enumeration}.

\subsection{Determinant form of TASEP weight with general $\alpha$, $\beta$}
\label{sec:AlphaBetaDeterminant}

Away from permutations and returning to the TASEP, Mandelshtam has generalised the determinant form \eqref{eq:PathDeterminantFormulaX} of a TASEP configuration weight for arbitrary $\alpha$, $\beta$ (Corollary 5.2 in \cite{mandelshtam2015determinantal}, modified to be consistent with notation used here):
\begin{equation}
\label{eq:MandelshtamDeterminant}
\fl \mathcal{W}(\mathcal{C}) = \frac{\det \;{M}}{\alpha^Q\beta^P}
\end{equation}
where the entries of $M$
\begin{eqnarray}
\fl M_{nm} = \beta^{m-n}\alpha^{-x_{n-1}} \Bigg\{\alpha ^{x_{m}} \left[\binom{Q-x_{m}}{m-n}+\beta \binom{Q-x_{m}}{m-n+1}\right] \label{eq:MandelshtamDeterminant2} \\
\qquad \qquad \qquad +\alpha ^{x_{m-1}} \sum _{l=0}^{x_{m}-x_{m-1}-1} \alpha ^l \left[\binom{Q-x_{m-1}-l}{m-n-1}+\beta 
\binom{Q-x_{m-1}-l}{m-n}\right]\Bigg\} \nonumber
\end{eqnarray}
with $n$, $m = 1, \dots P$, and the $x_n$, $x_m$ are the coordinates associated to an ASEP configuration in Section~\ref{sec:PathDominanceMapping}.
\begin{figure}[!t]
	\centering
	\includegraphics[scale=0.7]{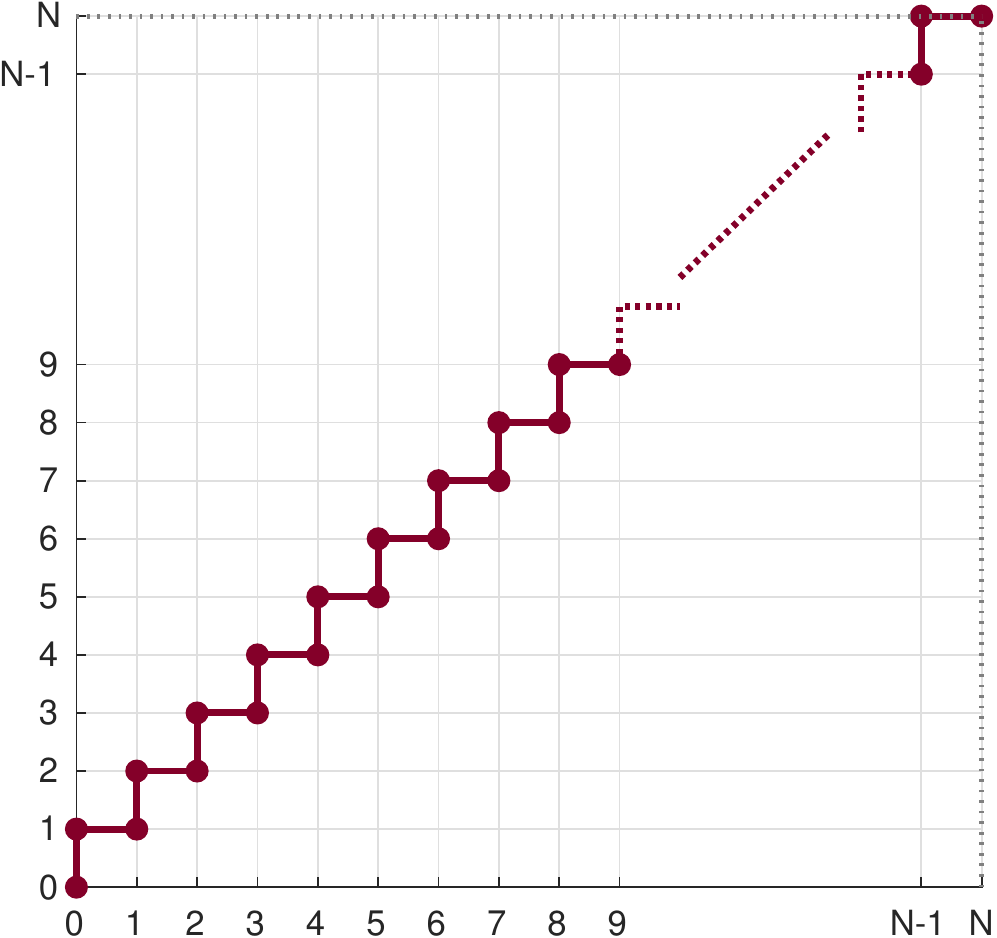}
	\caption{Partition function $(D+E)^N$ expressed as a `staircase' path.}
	\label{fig:PartitionFunctionStaircase}
\end{figure}

Using this formula, we are able to write down an expression for the TASEP partition function $Z_N$ in the form of  a  determinant. As far as we are aware the expression we now derive has not previously appeared in the literature. Given that
\begin{equation}
(D+E)^N = (DE)^N = \underbrace{(DE)\dots(DE)}_{N}\;,
\end{equation}
the partition function is the weight of a single `staircase' path of length $2N$ (see Figure \ref{fig:PartitionFunctionStaircase}). For this path, $x_m = m$, and Mandelshtam's formula \eqref{eq:MandelshtamDeterminant}, \eqref{eq:MandelshtamDeterminant2} eventually reduces to
\begin{eqnarray}
\label{eq:DeterminantPartitionFunction}
\fl Z_N = \mathrm{det} \; M \\
\fl M_{nm} = \binom{m-1}{n-m}\left(\frac{1}{\alpha}+\frac{1}{\beta}\right) + \binom{m-1}{n-m-1}\frac{1}{\alpha\beta} + \binom{m-1}{n-m+1} \\
\fl = \left(
\begin{array}{cccccc}
\left[\frac{1}{\alpha }+\frac{1}{\beta }\right] & 1 & \cdot & \cdot & \cdots \\
\frac{1}{\alpha \beta } & 1+\left[\frac{1}{\alpha }+\frac{1}{\beta }\right] & 1 &
\cdot & \cdots \\ \cdot & \left[\frac{1}{\alpha }+\frac{1}{\beta }\right] + \frac{1}{\alpha \beta }&
2 + \left[\frac{1}{\alpha }+\frac{1}{\beta }\right] & 1 & \cdots \\
\cdot & \frac{1}{\alpha \beta } & 1 + 2\left[\frac{1}{\alpha }+\frac{1}{\beta }\right] + \frac{1}{\alpha \beta
} & 3 + \left[\frac{1}{\alpha
}+\frac{1}{\beta }\right] & \cdots \\
\cdot & \cdot &\left[\frac{1}{\alpha }+\frac{1}{\beta }\right] + \frac{2}{\alpha \beta } & 3 + 
3\left[\frac{1}{\alpha }+\frac{1}{\beta }\right] + \frac{1}{\alpha \beta }& \cdots\\
\vdots & \vdots & \vdots & \vdots & \ddots \\
\end{array}
\right)_{N\times N}
\end{eqnarray}
where we see rows of Pascal's triangle in the coefficients of $1$, $\left[1/\alpha + 1/\beta\right]$, $1/\alpha\beta$ when reading down columns of $M$. We verify in \ref{sec:AppendixDeterminantGeneratingFunction} that \eqref{eq:DeterminantPartitionFunction} and the partition function are equivalent, through generating functions.

\subsection{$\alpha$, $\beta$ generalisation of path dominance problem}
\begin{figure}[!t]
	\centering
	\includegraphics{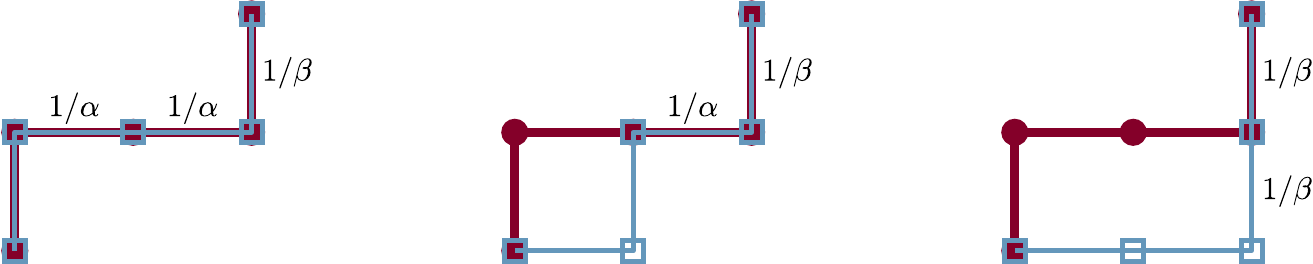}
	\caption{The weight of the path $\mathcal{T} = (\uparrow,\rightarrow,\rightarrow,\uparrow)$, corresponding to the TASEP configuration $\mathcal{C} = (1,0,0,1)$. Both have weight $\mathcal{W}(\mathcal{T}) = 1/\alpha^2\beta+1/\alpha\beta+1/\beta^2 = \bra{W}DEED\ket{V}$.}
	\label{fig:WeightedDominatedPaths}
\end{figure}

Following on from this determinant formula, there is a straightforward generalisation to $\alpha$, $\beta$ in the dominated path interpretation of TASEP weights. In the context of the original reference \cite{mandelshtam2015determinantal} these are referred to as ``weighted Catalan paths'', which translate into our formalism as follows: each dominated path has an associated weight $(1/\alpha)^p (1/\beta)^q$, where $p$ is the number of horizontal steps where both paths run together, and $q$ is the number of `up' steps the dominated path takes at the end of the walk. See Figure \ref{fig:WeightedDominatedPaths} for an example.

\subsection{General $\alpha$, $\beta$, $q$}

We have arrived at the most general case of general $\alpha$, $\beta$, $q$. We shall discuss a natural interpretation in terms of bicoloured Motzkin paths that arises from an explicit matrix representation. Otherwise, the most notable progress here has been by Corteel and Williams \cite{corteel2007tableaux}, who derive a generalised version of the path representation of configurations in Section~\ref{sec:PathDominanceMapping}, termed \emph{permutation tableaux}.

At this level of generality, there are few new physical insights that have been made other than establishment of the mapping.

\subsubsection{Weighted bicoloured Motzkin paths.}
\label{sec:WBCMPs}
In this context, the natural explicit representation to use is \cite{blythe2007nonequilibrium}
\begin{eqnarray}
\label{eq:AltExplicitMatrixRep1}
\fl D &= \frac{1}{1-q}\left( \begin{array}{ccccc}
1+b & \sqrt{c_0} & \cdot & \cdot & \cdots \\
\cdot & 1+bq & \sqrt{c_1} & \cdot & \cdots \\
\cdot & \cdot & 1+bq^2 & \sqrt{c_2} & \cdots \\
\cdot & \cdot & \cdot & 1+bq^3 & \cdots \\
\vdots & \vdots & \vdots & \vdots & \ddots \end{array} \right) \\
\fl E &= \frac{1}{1-q}\left( \begin{array}{ccccc}
1+a & \cdot & \cdot & \cdot & \cdots \\
\sqrt{c_0} & 1+aq & \cdot & \cdot & \cdots \\
\cdot & \sqrt{c_1} & 1+aq^2 & \cdot & \cdots \\
\cdot & \cdot & \sqrt{c_2} & 1+aq^3 & \cdots \\
\vdots & \vdots & \vdots & \vdots & \ddots \end{array} \right) \label{eq:AltExplicitMatrixRep2} \\
\fl \bra{W} &= (1,0,0,\cdots)\;, \qquad \ket{V} = (1,0,0,\cdots)^T \label{eq:AltExplicitMatrixRep4}
\end{eqnarray}
with $a$ and $b$ defined in \eqref{eq:abdefinition}, and $c_n = (1-q^{n+1})(1-abq^n)$. $D$ and $E$ then operate on a state ket $\ket{n}$
\begin{eqnarray}
D\ket{n} &= \frac{1}{1-q}\left(\sqrt{c_{n-1}}\ket{n-1} + (1+bq^n)\ket{n}\right) \\
E\ket{n} &= \frac{1}{1-q}\left(\sqrt{c_{n+1}}\ket{n+1} + (1+aq^n)\ket{n}\right)\;.
\end{eqnarray}
Note that this representation is distinct from (\ref{eq:ExplicitMatrixRep1}--\ref{eq:ExplicitMatrixRep4}).
\label{sec:AlphaBetaGeneral}
\begin{figure}[!t]
	\centering
	\includegraphics{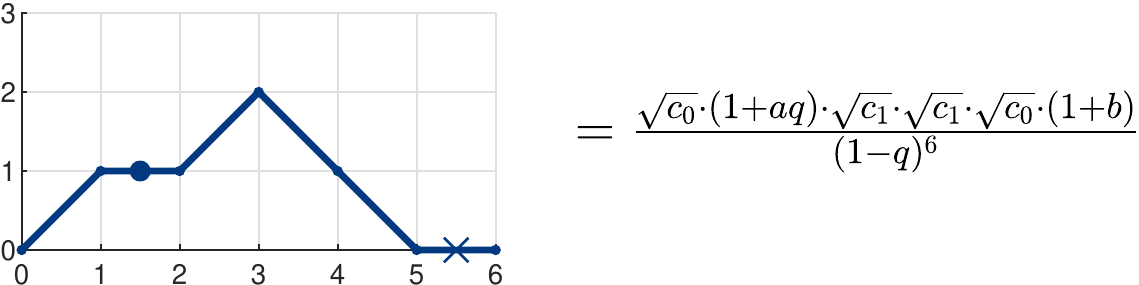}
	\caption{Weight of the Motzkin path $(\nearrow, \cdot, \nearrow, \searrow, \searrow, \times)$. This is one of many paths mapping to the configuration $\mathcal{C} = (1,1,1,0,0,0)$.}
	\label{fig:WeightedMotzkin}
\end{figure}

This representation lends a natural association of \emph{weights} on the bicoloured Motzkin paths \cite{brak2004asymmetric,brak2006combinatorial,blythe2009continued} (or equivalently, dominated paths). In Section~\ref{sec:MotzkinWalks} we inferred that the weight of a configuration $\mathcal{C}$ with sites $(j_1, j_2, \dots j_P)$ occupied is an enumeration of bicoloured Motzkin paths of length $N$, with steps $(j_1, j_2, \dots j_P)$ from $\{\nearrow,\,\cdot\,\}$, and the remaining steps from $\{\searrow, \times\}$. The same formalism applies here, except for each step we associate weights:
\begin{itemize}
	\item a $\nearrow$ from height $n$ to $(n+1)$ has weight $\sqrt{c_n}/(1-q)$,
	\item a $\searrow$ from height $(n+1)$ to $n$ has weight $\sqrt{c_n}/(1-q)$,
	\item a $\times$ at height $n$ has weight $(1+bq^n)/(1-q)$,
	\item a $\cdot$ at height $n$ has weight $(1+aq^n)/(1-q)$.
\end{itemize}
The weight of the path is then the product of these weights. See Figure \ref{fig:WeightedMotzkin} for an example. Note that $\nearrow$ and $\searrow$ always appear in pairs, eliminating the square root in factors of $\sqrt{c_n}$.

\subsection{Mapping between Motzkin paths and permutations for $\alpha=\beta=1$ and general $q$.}
\label{sec:WeightedPermsMapping}
We conclude this section by proposing a one-to-one mapping between a set of \emph{decorated} bicoloured Motzkin paths and a permutation of $(N+1)$ integers that applies for general $q$ when $\alpha=\beta=1$. These permutations have the integers $(j_1, j_2, \dots j_P)$ raised by the integers $(k_1, k_2, \dots k_P)$, and have the a weight $q^r$ as described in Section~\ref{sec:qMappingPermutationProblem}.

\begin{figure}
	\begin{center}
		\includegraphics{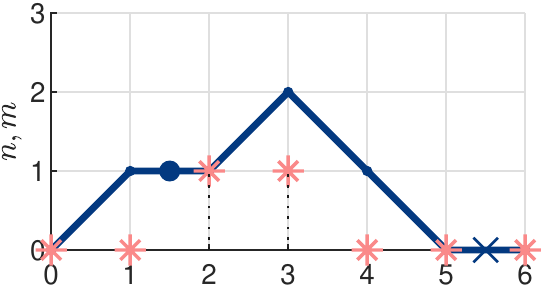}
	\end{center}
	\caption{\label{fig:DecoratedMotzkin} A decorated bicoloured Motzkin path of length $N=6$, with baubles (red, starred) at heights $m_i=(0,0,1,1,0,0,0)$. The weight of this path is $q^2$.}
\end{figure}

In addition to the usual up steps ($\nearrow$), down steps ($\searrow$) and horizontal steps of two colours ($\,\cdot\,$ and $\times$), the decorated bicoloured Motzkin path features at each horizontal position $i=0,1,\dots N$ a \emph{bauble} hanging at a height $m_i=0, 1, \dots n_i$, where $n_i$ is the height of the path at position $i$. An example of such a path is shown in Figure~\ref{fig:DecoratedMotzkin}. The weight of the path is $q^{r}$ where $r=\sum_{i=0}^{N} m_i$, i.e., the sum of the bauble heights.

Summing over all bauble positions for a given Motzkin path yields the weight of Section~\ref{sec:WBCMPs} after putting $\alpha=\beta=1$ (and therewith $a=b=-q$). To see this we note that each $\nearrow \dots \searrow$ pair between heights $n$ and $(n+1)$ contributes a weight $(1+q+\dots+q^n)(1+q+\dots+q^{n+1})$ while a $\times$ and a $\,\cdot\,$ at height $n$ each contribute a weight $(1+q+\dots+q^n)$.

Each possible decorated bicoloured Motzkin path can be translated into a permutation of the integers $(0, 1, \dots N)$ through an iterative procedure that we now describe. The basic idea is that the horizontal position at the start of each up step or horizontal step of type $\,\cdot\,$ determines the integers $j_\ell$ that are raised, and the horizontal position at the end of each down step or horizontal step of type $\,\cdot\,$ determines the integers $k_\ell$ that do the raising. The baubles determine the order that integers appear in the permutation. The algorithm ensures that the constraint that \emph{only} the integers $j_{\ell}$ are raised is never violated.

The iterative procedure begins with an empty string, and at step $i$ involves placing the integer $i$ into the string. At intermediate steps, the string may contain \emph{placeholder elements} which receive the integers $k_{\ell}$ that do the raising. For each successive integer $i=0,1,\dots N$, the iteration comprises two sub-steps:
\begin{enumerate}
	\renewcommand{\theenumi}{\alph{enumi}}
	\item\label{prevstep} If the previous segment (that connecting $i-1$ to $i$) is a down-step or a horizontal step of type $\,\cdot\,$, \emph{replace} the $(m_i+1)^{\rm th}$ placeholder from the left with the integer $i$. Otherwise, if $m_i>0$, insert integer $i$ \emph{after} the $m_i^{\rm th}$ placeholder. Otherwise, place integer $i$ at the start of the string. (Note: when $i=0$ the only option is for the string to become $0$.)
	\item\label{nextstep} If the next segment (that connecting $i$ to $i+1$) is an up-step or a horizontal step of type $\,\cdot\,$, \emph{insert} a placeholder element \emph{after} the integer $i$. (Note: when $i=N$, there is no next segment, and the algorithm terminates.)
\end{enumerate}
In Table~\ref{tab:Iteration} we illustrate this procedure with the example path and decoration shown in Figure~\ref{fig:DecoratedMotzkin}, thereby determining that it corresponds to the permutation $6\,1\,4\,3\,0\,2\,5$.

\begin{table}
	\begin{center}
		\begin{tabular}{l|ll|ll}
			$i$ & \multicolumn{2}{l|}{Sub-step (\ref{prevstep})}  & \multicolumn{2}{l}{Sub-step \eqref{nextstep}} \\\hline
			0 & $0$ & First step & $0{\circ}$ & Up step: add ${\circ}$ \\
			1 & $1\,0\,{\circ}$ & $m_1=0$: insert $1$ at start & $1\,{\circ}\,0\,{\circ}$ &  Horiz.\ $\,\cdot\,$ step: add ${\circ}$ \\
			2 & $1\,{\circ}\,0\,2$ & $m_2=1$: replace $2^{\rm nd}$ ${\circ}$ with $2$ & $1\,{\circ}\,0\,2\,{\circ}$ & Up step: add ${\circ}$ \\
			3 & $1\,{\circ}\,3\,0\,2{\circ}$ & $m_3=1$: insert $3$ after $1^{\rm st}$ ${\circ}$ & $1\,{\circ}\,3\,0\,2\,{\circ}$ & Down step: no change \\
			4 & $1\,4\,3\,0\,2\,{\circ}$ & $m_4=0$: replace $1^{\rm st}$ ${\circ}$ with $4$ & $1\,4\,3\,0\,2\,{\circ}$ & Down step: no change \\
			5 & $1\,4\,3\,0\,2\,5$ & $m_5=0$: replace $1^{\rm st}$ ${\circ}$ with $5$ & $1\,4\,3\,0\,2\,5$ &  Horiz.\ $\times$ step: no change \\
			6 & $6\,1\,4\,3\,0\,2\,5$ & $m_6=0$: insert $6$ at start & $6\,1\,4\,3\,0\,2\,5$ & Finished
		\end{tabular}
	\end{center}
	\caption{\label{tab:Iteration} Iteration of the two sub-steps (\ref{prevstep}) and (\ref{nextstep}) of the algorithm that translates a decorated bicoloured Motzkin path into a permutation. The symbol $\circ$ denotes a placeholder element.}
\end{table}

Since each integer is added to the string exactly once, we must end up with some permutation of the $(N+1)$ integers $(0, 1, \dots N)$. Sub-step (\ref{nextstep}) of the algorithm ensures that a placeholder is inserted immediately after any integer that must be raised. The first clause of sub-step (\ref{prevstep}) ensures that those integers that do the raising are inserted into the placeholders: these integers must necessarily be larger than those to the left of the placeholder, and so these are indeed raised as required. The remaining clauses of sub-step (\ref{prevstep}) insert the non-raising integers: these are always entered either at the start of the string (where no raising is possible) or after a placeholder element that will receive a higher integer at a later stage of the algorithm (and therefore also do not raise). Thus the resulting string has integers $(j_1, j_2 \ldots j_P)$ raised, and integers $(k_1, k_2, \dots k_P)$ doing the raising, as required. The height of the path keeps track of the number of placeholders and therefore the number of places where each successive integer can be placed while respecting the constraints. The position of the bauble determines how many placeholders lie to the left of the inserted integer, and therefore the power of $q$ that contributes to the weight. 

To show that the decorated paths map one-to-one to permutations, one needs to show that each permutation obtained from the above algorithm is distinct. For two paths of the same shape, but different bauble positions, this seems plausible, because the bauble position determines the position of each new integer $i$ relative to those already present in the string. The relative order of the first $i$ integers can not be changed by later insertions, so one expects each bauble configuration to map to a distinct permutation. Meanwhile, different permutations correspond to different sets of raised or raising integers, and therewith a partitioning of the permutations into distinct subsets. Finally, it is already established in \eqref{eq:SSEPPartitionFunctionab1} that $\lim_{q\to1} Z(\alpha,\beta,q) = (N+1)!$ for $\alpha=\beta=1$. Since this normalisation counts the total number of decorated bicoloured Motzkin paths, it would then follow that every possible permutation is represented by one of the paths. 

A formal proof of the proposition that there is a one-to-one mapping between decorated Motzkin paths and permutations would be welcome. In particular, it would demonstrate the one-to-many mapping from ASEP configurations to bicoloured Motzkin paths which themselves map one-to-many to permutations as was illustrated in Figure \ref{fig:DraftOverviewDiagram}. For each bicoloured Motzkin path, exactly one mapped permutation has weight $q^0 = 1$, that is, the decorated path with baubles at $m = (0,0,\dots 0)$), while all others have a positive power of $q$. This would connect the path dominance mapping of the TASEP ($q\to0$), and the permutation mapping of the SSEP ($q\to1$).

\subsubsection{Permutation and staircase tableaux.}\label{sec:perm}

Elsewhere, Corteel and Williams \cite{corteel2007tableaux,corteel2011tableaux,corteel2012formulae} have mapped the most general case of $\alpha$, $\beta$, $q$ to a problem in an area known as \emph{tableaux combinatorics}. We refer the reader to \cite{corteel2007tableaux} for the original work, and \cite{corteel2011tableaux, corteel2012formulae} for a more generalised case of \emph{staircase tableaux} that encodes two extra parameters $\gamma$, $\delta$ (so particles may also enter from the right, and leave from the left).

The details are beyond the scope of this work, but to sketch their approach the authors take ASEP configurations as paths drawn in Section~\ref{sec:PathDominanceMapping}, and construct a grid across the area the path bounds (a \emph{Young diagram}). Each entry of this grid can take a value of $\alpha$, $\beta$, $q$ or $1$ (or a generalised hop-right rate $u$), with a set of rules as to which values can go where. The weight of this \emph{permutation tableaux} is then the product of all of the entries. For a given ASEP configuration a set of these permutation tableaux can be drawn, and the weight of the configuration is the sum of weights of these tableaux. Further combinatorial interpretations of the PASEP partition function have been obtained by Josuat-Verg\`es \cite{JV2011tableaux}: as the generating function for weighted Motzkin paths known as Laguerre histories, and as the generating function for permutations with respect to maxima and minima and other features.

\section{State space measures of the TASEP}
\label{sec:LambdaWeightEnumeration}

We now shift our focus to functions that involve \emph{all} weights of the TASEP. This is a different type of problem to the mappings discussed up to now, as we are now calculating measures of the state space as a whole, instead of individual configurations. We will still use the random walk interpretation offered by the explicit matrix representations, but now in higher dimensions.

By exactly solving the TASEP and its partition function, we can calculate steady-state quantities such as the particle current $J$, the steady-state density profile $\langle\tau_i\rangle$ and higher-order correlations between different sites $\langle \tau_i\tau_j\rangle$ \cite{derrida1993exactcorr}. However, in the absence of equilibrium statistical mechanics, how do we probe the finer details of the probability distribution?

\subsection{R\'enyi entropy}

To this end, we introduce the \emph{R\'enyi entropy}, which is defined \cite{Renyi1961}
\begin{equation}
\label{eq:RenyiEntropy}
H_\lambda = \frac{1}{1-\lambda}\log\left(\sum_{\mathcal{C}} \mathcal{P}(\mathcal{C})^\lambda\right) = \frac{1}{1-\lambda} \log \left(\sum_{\mathcal{C}} \left[\frac{\mathcal{W}(\mathcal{C})}{\sum_{\mathcal{C}'} \mathcal{W}(\mathcal{C}')}\right]^\lambda \right)
\end{equation}
as a measure of a full probability distribution. $\lambda$ is a nonnegative number. To provide an interpretation of $H_\lambda$, consider
\begin{equation}
\label{eq:EffectiveNumber}
\e^{H_\lambda} = \left[\sum_{\mathcal{C}} \mathcal{P}(\mathcal{C})^\lambda\right]^\frac{1}{1-\lambda} \;.
\end{equation}
A system with $L$ equally likely configurations has $\e^{H_\lambda} = L$. At the other extreme, the same system with a single configuration with probability one has $\e^{H_\lambda} = 1$. Between these extremes, then, $\e^{H_\lambda}$ measures an \emph{effective number} of participating configurations. By increasing $\lambda$, the measure places more weight on the higher $\mathcal{P}$ values, as the lower ones are exponentially suppressed. At the extreme, $\e^{H_0}$ is the number of configurations with nonzero probability.

The $\lambda \to 1$ limit recovers the familiar Shannon entropy:
\begin{equation}
\label{eq:ShannonLimit}
\lim_{\lambda \to 1}H_\lambda = 
\lim_{\lambda \to 1} \frac{\log{\sum_{\mathcal{C}} \mathcal{P}(\mathcal{C}) {\rm e}^{(\lambda-1) \log \mathcal{P}(\mathcal{C})}}}{1-\lambda} = -\sum_{\mathcal{C}} \mathcal{P}(\mathcal{C})\log \mathcal{P}(\mathcal{C}) \;.
\end{equation}
For all $\lambda$ the R\'enyi entropy \eqref{eq:RenyiEntropy} is exactly calculable for any equilibrium system with a known partition function, as (aside from the special $\lambda = 1$ case) a ratio of partition functions at different temperatures \cite{Baez2011}. This includes the one-transit walk in Section~\ref{sec:OneTransitWalk}, which is an equilibrium system.

We have seen from these combinatorial mappings that weights of the TASEP do not take such a convenient exponential form (e.g. Equation \eqref{eq:ExamplePASEPWeight}) and are summations that can not generally be factorised. Furthermore, even though the ASEP shares a partition function with an equilibrium system, any sum of weights to a power $\lambda$ depends on how the $C_{N+1}$ equilibrium configurations map onto $2^N$ nonequilibrium configurations---in other words, the partitioning of the partition function.

\subsubsection{Enumeration of weights raised to a power.}

Given the form of the R\'enyi entropy in \eqref{eq:RenyiEntropy}, we are interested in calculating
\begin{equation}
\label{eq:SumWeightsLambda}
\sum_{\mathcal{C}} \mathcal{W}\left(\mathcal{C}\right)^{\lambda}
\end{equation}
for the TASEP, for different values of $\lambda$. As an introductory example, we first calculate the partition function (for general $\alpha$, $\beta$), in terms of weighted bicoloured Motzkin paths. This is the straightforward $\lambda = 1$ case of \eqref{eq:SumWeightsLambda}.

We then show that the problem of \eqref{eq:SumWeightsLambda} is --- for integer $\lambda$ --- equivalent to a lattice enumeration problem in $\lambda$ dimensions (the partition function being the one-dimensional problem). The $\lambda = 2$ case was previously solved by the authors in \cite{wood2017renyi}.

\subsection{$\lambda = 1$; sum of weights}

\label{sec:WeightEnumeration}
With $q=0$, the explicit representation of the matrices $D$, $E$ and vectors $\bra{W}$, $\ket{V}$ in (\ref{eq:ExplicitMatrixRep1}--\ref{eq:ExplicitMatrixRep4}) reduce to
\begin{eqnarray} 
\label{eq:ExplicitDERepLadder1}
D = \textbf{1} + g & \qquad E = \textbf{1} + g^\dagger \nonumber\\
D\ket{k}=\ket{k}+\ket{k-1} & \qquad E\ket{k}=\ket{k}+\ket{k+1} \nonumber\\
\bra{k} D=\bra{k}+\bra{k+1} & \qquad \bra{k} E=\bra{k}+\bra{k-1} \;. 
\end{eqnarray}
Then the boundary vectors
\begin{equation}
\label{eq:ExplicitDERepLadder2}
\fl \bra{W} = \sqrt{1-ab}\;(1, a, a^2, a^3, \dots)\;, \qquad \ket{V} = \sqrt{1-ab}\;(1, b, b^2, b^3, \dots)^T\;,
\end{equation}
recalling $a = (1-\alpha)/\alpha$, $b = (1-\beta)/\beta$ \eqref{eq:abdefinition}. From this, the partition functiion $Z_N$ is then 
\begin{eqnarray}
Z_N &= \bra{W}(D+E)^N\ket{V} \\
& = (1-ab)\sum_{i\geq 0}\sum_{k \geq 0} a^ib^k\bra{i}\left(g + g^\dagger + 2\right)^N\ket{k} \label{eq:PartitionFunctionFunnyWalk}
\end{eqnarray}
after writing the scalar product explicitly. The RHS of \eqref{eq:PartitionFunctionFunnyWalk} takes the form of a \emph{generating function} in $a$, $b$ of the quantity $\bra{i}\left(g + g^\dagger + 2\right)^N\ket{k}$. Given the binomial expansion of $\left(g + g^\dagger + 2\right)^N$ and that $g\ket{0} = 0$, this quantity is the number of bicoloured Motzkin paths of length $N$ between coordinates $i$ and $k$.

The partition function therefore follows from the path enumeration problem $\bra{i}(g+g^\dagger+2)^N\ket{k}$. One way of solving this is by generating functions. We present a neat example of this by Depken \cite{depken2003models}: first, the generating function is written
\begin{equation}
\mathcal{Z}(z) = \sum_{N\geq0} \bra{W}(D+E)^N\ket{V} z^N = \bra{W}\frac{1}{1-z(D+E)}\ket{V} \;.
\end{equation}
Now using relation \eqref{eq:ReductionRelations1TASEP}, we find
\begin{equation}
(1-\eta D)(1-\eta E) = 1-\eta(D+E) + \eta^2DE = 1-\eta(1-\eta)(D+E) \;.
\end{equation}
Taking $z = \eta(1-\eta) \Rightarrow \eta(z) = \hf(1-\sqrt{1-4z})$, this gives
\begin{equation}
\mathcal{Z}(z) = \bra{W}\frac{1}{1-\eta(z)E}\frac{1}{1-\eta(z)D}\ket{V} = \frac{1}{1-\frac{\eta(z)}{\alpha}}\frac{1}{1-\frac{\eta(z)}{\beta}}
\end{equation}
taking the negative root of $\eta(z)$ to ensure $\mathcal{Z}(0) = 1$.

We should expect this problem to simplify in the case $\alpha = \beta = 1$ ($a = b = 0$). Indeed this is the case, and \eqref{eq:PartitionFunctionFunnyWalk} reduces to 
\begin{equation}
Z_N = \bra{0}(g+g^\dagger+2)^N\ket{0}
\end{equation}
which is the number of such walks that start and end at zero. As discussed in Section~\ref{sec:DyckMotzkin} and \ref{sec:AppendixCatalanDerivation} this is of course the Catalan number $C_{N+1}$.

With this, we return to the central problem of \eqref{eq:SumWeightsLambda} and R\'enyi entropies. Configurations of the TASEP map to $C_{N+1}$ paths. The partition function $Z_N$ is then a summation of this larger state space. The summation of TASEP weights each raised to a \emph{power}, however, is a much more complex problem as it depends specifically on which paths map to which TASEP configurations. We first take the sum of \emph{squared} weights.

\subsection{$\lambda = 2$: sum of squared TASEP weights}

\begin{figure}[!t]
\centering 
\includegraphics{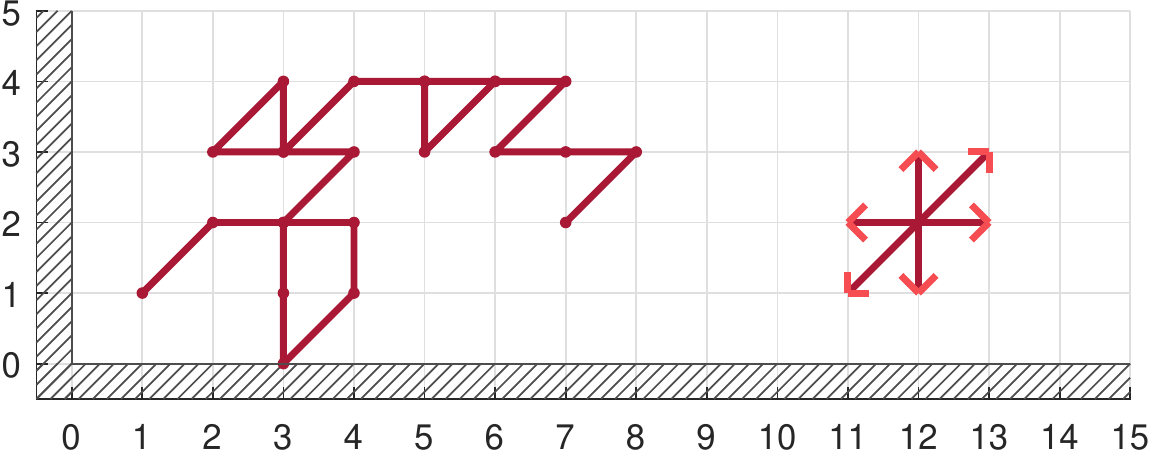}
	\caption{Example of a 2D walk comprising the steps $\{\uparrow, \downarrow, \rightarrow, \leftarrow, \swarrow, \nearrow \}$ from $(i,j)=(1,1)$ to $(k,l)=(7,2)$. The walk must remain in the upper quadrant, but may touch and move along the boundary.}
	\label{fig:2DFunnyWalk}
\end{figure}

The sum of squared weights of the ASEP is compactly written in \emph{tensor} product formalism \cite{wood2017renyi}:
\begin{equation}
\label{eq:SumSquaresTensorProduct}
\sum_{\mathcal{C}} \mathcal{W}(\mathcal{C})^2 = \bra{W}\otimes\bra{W}\left(D\otimes D + E\otimes E \right)^N\ket{V}\otimes\ket{V} \;,
\end{equation}
where $A \otimes B$ is the tensor product of $A$ and $B$. Taking the example of $N=2$
\begin{eqnarray}
\fl \bra{W}\otimes \bra{W}(D\otimes D +E\otimes E)^2\ket{V} \otimes \ket{V} \nonumber \\
\fl = \bra{W}\otimes \bra{W}(DD\otimes DD+ EE\otimes EE + ED\otimes ED +DE\otimes DE )\ket{V} \otimes \ket{V} \\
\fl = \left(\bra{W}DD\ket{V}\right)^2 + \left(\bra{W}EE\ket{V}\right)^2 + \left(\bra{W}ED\ket{V}\right)^2 +\left(\bra{W}DE\ket{V}\right)^2 \;.
\end{eqnarray}
We write the explicit representation in \eqref{eq:ExplicitDERepLadder1}, \eqref{eq:ExplicitDERepLadder2}, now in two dimensions:
\begin{eqnarray}
D\otimes D &= (1 + g_1 + g_2 + g_1g_2) \\
E\otimes E &= (1 + g^\dagger_1 + g^\dagger_2 + g^\dagger_1g^\dagger_2)
\end{eqnarray}
defining the ladder operators $g_1$, $g_2$:
\begin{eqnarray}
g_1 \ket{k}\otimes\ket{l} &= \ket{k-1}\otimes\ket{l}\;, \qquad 
g_1^\dagger \ket{k}\otimes\ket{l} &= \ket{k+1}\otimes\ket{l}\;, \\
g_2 \ket{k}\otimes\ket{l} &= \ket{k}\otimes\ket{l-1}\;, \qquad g_2^\dagger \ket{k}\otimes\ket{l} &= \ket{k}\otimes\ket{l+1} \;.
\end{eqnarray} 
We can now explicitly write the tensor product \eqref{eq:SumSquaresTensorProduct} as
\begin{eqnarray}
\label{eq:SumSquaresFunnyWalk}
\fl \frac{\sum_{\mathcal{C}} \mathcal{W}({\cal C})^2}{(1-ab)^2} \nonumber \\
\fl = \sum_{i \geq 0}\sum_{j\geq 0}\sum_{k\geq 0}\sum_{l\geq 0} a^ia^jb^kb^l\bra{i}\otimes\bra{j}(g_{1}+g_{2}+g_{1}g_{2}+g^\dagger_{1}+g^\dagger_{2}+g_1^\dagger g_2^\dagger+2)^N\ket{k}\otimes\ket{l} 
\end{eqnarray}
and again the RHS resembles a generating function in $a$, $a$, $b$, $b$ for the quantity $\bra{i}\otimes\bra{j}(g_{1}+g_{2}+g_{1}g_{2}+g^\dagger_{1}+g^\dagger_{2}+g_1^\dagger g_2^\dagger+2)^N\ket{k}\otimes\ket{l}$. This is a \emph{two} dimensional lattice walk; the six ladder terms can be interpreted as the step set $\{\uparrow, \downarrow, \rightarrow, \leftarrow, \swarrow, \nearrow \}$, alongside two non-movement steps `$\times$', `$\cdot$'. \eqref{eq:SumSquaresFunnyWalk} is thus a generating function of the number of walks in the upper quadrant of length $N$ between $(i,j)$ and $(k,l)$ from the step set $\{\uparrow, \downarrow, \rightarrow, \leftarrow, \swarrow, \nearrow, \times, \cdot \}$. See Figure \ref{fig:2DFunnyWalk} for such a walk. In practice the non-movement steps are easily integrated out.

In an earlier work \cite{wood2017renyi} we calculated the generating function \emph{of} \eqref{eq:SumSquaresFunnyWalk}
\begin{equation}
\label{eq:SumSquaresGeneratingFunction}
\sum_{N\geq 0} z^N \sum_{\mathcal{C}} \mathcal{W}(\mathcal{C})^2 \;.
\end{equation}
The enumeration of these \emph{upper quadrant walks} for a general step set is is a well-researched topic, and a technique known as the \emph{kernel} method finds the generating function for most walks \cite{bousquet2010walks}. The step set we are dealt with here proves to be one of the more stubborn, and is solved by a more involved \emph{obstinate} kernel method. While the details of the calculation are beyond the scope of this review, the symmetry of the six-step walk is exploited in order to solve what turns out to be a two-parameter recursion relation of the generating function \cite{bousquet2010walks}.

Having found an explicit form of \eqref{eq:SumSquaresGeneratingFunction}, the asymptotic scaling of the sum of squared weights emerges, with a different scaling for the three phases. After normalising,
\begin{equation}
\sum_{\mathcal{C}} \mathcal{P}(\mathcal{C})^2 \sim
\cases{ f(\alpha,\beta)\left(\alpha^2+(1-\alpha)^2\right)^N & LD $\alpha < \beta, \alpha < \hf$ \\
	f(\beta,\alpha)\left(\beta^2+(1-\beta)^2\right)^N & HD $\beta < \alpha, \beta < \hf $ \\
	h(\alpha,\beta)2^{-N} N^{\hf} & MC $\alpha > \hf, \beta > \hf$} \;.
\end{equation}
We briefly mention that the scaling in this maximal current phase implies that the effective number scales as $2^N/\sqrt{N}$, which is the same asymptotic scaling as the binomial coefficient $\left(N\atop N/2\right)$. Given the MC phase has a density of $\tau_i \approx 1/2$ in the bulk, the effective number is in turn proportional to the number of half-filled configurations.

\subsubsection{$\alpha = \beta = 1$ simplification; direct solution.} 
\label{sec:SumSquaresDominance}

Setting $\alpha = \beta = 1$ ($a=b=0$), \eqref{eq:SumSquaresFunnyWalk} is the enumeration of walks that start and end at the origin only. Here, the generating function is succinct \cite{wood2017renyi}:
\begin{eqnarray}
\fl \sum_{N\geq 0} z^N \sum_{\mathcal{C}} \mathcal{W}(\mathcal{C})^2 = \frac{1}{4z^2}\bigg[3\sqrt{2z}\sqrt{1-2z-\sqrt{1-8z}} \\ \qquad\qquad\qquad\qquad +\sqrt{2(1+z)}\sqrt{1-2z+\sqrt{1-8z}}-4z-2\bigg] \nonumber \\
\fl \qquad \qquad \quad \qquad = 1+2z+7z^2+30z^4+146z^5+772z^6 \dots \;.
\end{eqnarray}
The coefficients of this series expansion are (sequence A196148 in the OEIS \cite{oeisSumSquares})
\begin{equation}
\label{eq:SumSquaresAlphaBeta1}
\sum_{\mathcal{C}} \mathcal{W}(\mathcal{C})^2 = \sum_{P=0}^N \frac{(2N+1)!(N+1)!}{(2P + 1)!(2N - 2P + 1)!(P + 1)!(N - P + 1)!}
\end{equation}
and the sum of squared weights for configurations with $P$ particles is \eqref{eq:SumSquaresAlphaBeta1} with the summation dropped (sequence A111910 \cite{oeisSumSquaresP}). This can be proven by demonstrating that this number sequence and \eqref{eq:SumSquaresFunnyWalk} have the same generating function. These numbers take a similar form to Narayana numbers \eqref{eq:NarayanaNumber}.

\begin{figure}[!t]
	\includegraphics{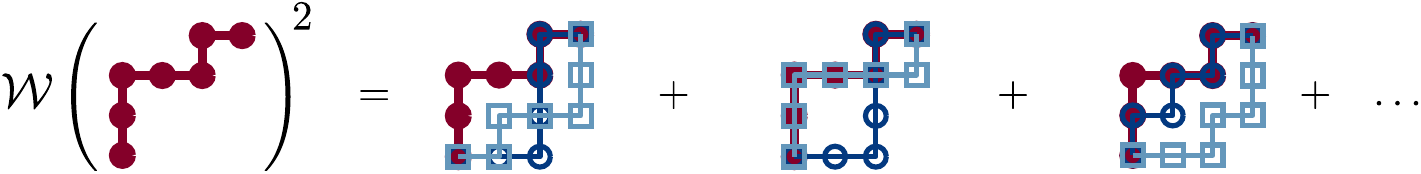}
	\caption{Graphical representation of the squared weight of a path in the path dominance formalism.}
	\label{fig:SquaredDominanceWeights}
\end{figure}

It is interesting to note that a path dominance problem closely related to the sum of square weights was solved directly by Kreweras and Niederhausen in \cite{kreweras1981solution} outside of the context of the TASEP. The equivalent problem is the enumeration of \emph{triples} of dominated paths: paths that can be drawn where one path dominates the other two (Figure \ref{fig:SquaredDominanceWeights}). 

\subsection{Higher orders}
This path enumeration approach to sums of TASEP weights can be generalised to arbitrary integer power. The sum of weights to the $\lambda^\mathrm{th}$ power can be written
\begin{equation}
\sum_{\mathcal{C}} \mathcal{W}(\mathcal{C})^\lambda = \bra{W}^{\otimes \lambda}(D^{\otimes \lambda}+E^{\otimes \lambda})\ket{V}^{\otimes \lambda} \;,
\end{equation}
where 
\begin{equation}
A^{\otimes \lambda} = \underbrace{A\otimes \dots \otimes A}_{\lambda} \;.
\end{equation}
Given the explicit representation of $D$ and $E$, this then reduces to a problem of enumerating the $\lambda$-dimension walks in the upper \emph{orthant}, from the $2^{\lambda+1}$ steps arising from
\begin{equation}
\prod_{q=1}^\lambda(1+g_{q}) + \prod_{q=1}^\lambda(1+g^\dagger_{q}) \;.
\end{equation}
Even in two dimensions, the step set $\{\uparrow, \downarrow, \rightarrow, \leftarrow, \swarrow, \nearrow \}$ proved one of the more challenging step sets to solve. The enumeration of $\lambda = 3$ octant walks is a current area of research \cite{bostan20163,bacher2015continued}, but for this particular classification of walk in $\lambda = 3$ or higher, no analytical techniques are known.

This problem can be equivalently posed as a path dominance problem, extending the $\lambda = 2$ case in \cite{kreweras1981solution}. The sum of TASEP weights to the $\lambda^\mathrm{th}$ power is the number of distinct $(\lambda+1)$-tuples of length-$N$ paths that can be drawn, where one path dominates the $\lambda$ others.

\section{Multispecies models and further combinatorial connections}
\label{sec:multi}
So far we have considered in detail the matrix product formulation of the stationary state of the open ASEP and its combinatorial ramifications. As we have seen, this formulation leads to remarkably  far-reaching connections between matrix products, lattice paths and weighted permutations.

In this section we briefly give an outlook on how other exclusion processes also have matrix product solutions, and how matrix products may be generalised. The matrix product solution may be applied to a variety of multispecies problems as reviewed in \cite{blythe2007nonequilibrium}. Here we focus on recent progress on the family of models that involve a hierarchy of particle species: first-class particles, second-class particles and so on. 

Our aim here is to  highlight further connections between matrix product states and combinatorial constructions and indeed queueing theory.

\subsection{Second-class particles}
One of the first generalisations of the matrix product state of the open ASEP was to a system of first and second-class particles on a ring \cite{DJLS1993}. Here we first consider the partially asymmetric case. In this system first-class (normal) particles hop to the right with rate $1$ and to the left with rate $q$. They also \emph{overtake} second-class particles to the right with rate $1$ and to the left with rate $q$. Second-class particles hop to the right with rate $1$ and to the left with rate $q$. Thus first-class particles effectively view second-class particles as holes and from the point of view of holes, second-class particles behave in the same way as first-class particles. The dynamics may be summarised as
\begin{eqnarray}
1\;0 \to 0\;1 \quad\mbox{with rate} \quad 1\;,\\
0\;1 \to 1\;0 \quad\mbox{with rate} \quad q\;,\\
2\;0 \to 0\;2 \quad\mbox{with rate} \quad 1\;,\\
0\;2 \to 2\;0 \quad\mbox{with rate} \quad q\;,\\
1\;2 \to 2\;1 \quad\mbox{with rate} \quad 1\;,\\
2\;1 \to 1\;2 \quad\mbox{with rate} \quad q\;. \label{eq:multi}
\end{eqnarray}
We generalise our earlier single-species notation, and use the variable $\tau_i =0,1,2$ which now implies that site $i$ is empty, contains a first-class particle or contains a second-class particle, respectively. Let us denote by ${\cal C}=(\tau_1,\dots \tau_N)$, a configuration of the system.

The  matrix product solution of the stationary state was formulated by Derrida, Janowsky, Lebowitz and Speer \cite{DJLS1993}. 
In the matrix product formulation on a periodic lattice we now express the weight as a trace of a product of matrices $X_{\tau_i}$
\begin{equation}
\mathcal{W}({\cal C})= \mbox{Tr} \left[
  \prod_{i=1}^N X_{\tau_i} \right]\;.
\label{trace}
\end{equation}
The periodicity of the lattice is reflected in the periodicity of the trace operation on a product of matrices. The normalisation $Z=Z(N,P_1,P_2)$ now depends on the numbers $P_1$, $P_2$ of first and second-class particles respectively. The matrices $X_{\tau_i}$ are given by
\begin{equation}
X_0= E\;,\quad X_1=D\;,\quad X_2= A\;,
\end{equation}
that is, as before, we write a matrix $D$ or a matrix $E$ when the site contains a first-class particle or hole respectively, and if the site contains a second-class particle we write a matrix $A$.

The matrices $D$, $E$, $A$ obey the algebraic rules
\begin{eqnarray}
DE-qED &=&D + E \label{DE}\\
DA-qAD &=& A\label{DA}\\
AE -qAE &=& A \; . \label{AE}
\end{eqnarray}
The first equation (\ref{DE}) is as before. The new element is the matrix $A$.

A convenient representation is given by choosing $D$ and $E$ as before in \eqref{eq:ExplicitMatrixRep1} and \eqref{eq:ExplicitMatrixRep2}, and $A$ as the diagonal matrix
\begin{eqnarray}
\label{Amat}
A &= \frac{1}{1-q} \left( \begin{array}{ccccc}
1 & 0 & 0 & 0 & \cdots \\
0 & q & 0& 0 & \cdots \\
0 &  0 & q^2 & 0 & \cdots \\
0 & 0 & 0 & q^3 & \cdots \\
\vdots & \vdots & \vdots & \vdots & \ddots \end{array} \right) \;.
\end{eqnarray}

In the totally asymmetric case $q=0$ \eqref{Amat} reduces to
\begin{equation}
A = |0\rangle \langle 0|\;.
\label{eq:A}
\end{equation}
Note that this projector obeys $A^2$ = $A$ and due to the form of $A$, \eqref{trace} reduces to
\begin{equation}
\mathcal{W}({\cal C}) = \prod_{j=1}^{P_2}
\omega(B_j)
\label{P2mat2}
\end{equation}
where $B_j$ is a binary string and $\omega(B_j)$ is given by
\begin{equation}
\omega(B_j) = \langle0 | \prod_{i = 1}^\ell X_{\tau_i} | 0 \rangle
\label{Bstring}
\end{equation}
where $X_{\tau_i}$ is either a $D$ or $E$ according to whether $\tau_i$ is 1 or 0, $\ell$ is the length of the binary string $B_j$ and $i$ labels the entries in that string. Thus the weight of a two species configuration factorises around the second-class particles and reduces to a product of weights \eqref{Bstring} of open TASEP configurations which can be computed as before using the reduction rule \eqref{DE}. 

The matrices $D$, $E$ and $A$ with $A$ given by (\ref{Amat}) may also be used in the open boundary case under certain conditions on the boundary rates for entrance and exit of particles \cite{Evans1995}. In particular the case of `semi-permeable' boundaries where the second-class particles do not exit or enter has been widely studied \cite{Arita2006,Ayyer2009,Cantini2017,Aas2019}. More general open boundary versions  require generalisation of the matrices $D$, $E$, $A$ to tensor operators and certain conditions on boundary rates \cite{Crampe2015,Crampe2016} which relate to  Zamolodchikov-Faddeev and Ghoshal-Zamolodchikov relations in integrable systems \cite{Crampe2014}.

\subsection{Combinatorial connections}
In the case of second and first-class paricles with $q=0$ on the ring,
which we refer to as the 2-TASEP, Ferrari, Fontes and Kohayakawa
\cite{Ferrari1994} were able to use the factorisation property described above to give a complete probabilistic
description of the stationary measure.  Subsequently, Angel  \cite{Angel2006} was able to make a two-line construction, on line 1 of which there are $P_1$ particles distributed randomly and similarly on line 2 there are $P_1+P_2$ particles distributed randomly.  Angel  \cite{Angel2006}  showed that uniformly sampling two-line configurations, then generating the associated 2-TASEP configurations, samples 2-TASEP configurations according to their stationary measure. Thus it follows that the matrix representation of the stationary state furnishes a way of counting the two-line configurations which correspond to a 2-TASEP configuration. Angel also proved that the uniform measure on the two-line configurations is stationary under the dynamics.

Ferrari and Martin \cite{Ferrari2007} were  able to define a two-line process and show that the stationary distribution for the two-line process is the uniform distribution.  Specifically one can associate to each transition out of any two-line configuration a transition into that configuration such that there is bijection in the transition rates i.e.\ each member of the collection of transition events out of a configuration is paired with exactly one of the collection of transition events that lead in to that configuration. Then it follows that the stationary state has uniform probability for all allowed configurations of the two-line system. This is essentially the same argument that was used to show a uniform distribution over complete configurations for the TASEP with open boundaries described in Section~\ref{sec:completeconfigs} (see also \ref{sec:AppendixSimplifiedTwoRow}).

Furthermore, in \cite{Ferrari2007}, Ferrari and Martin showed that the two-line construction could be interpreted as a queueing system. The discrete `time' of the queueing process corresponds to lattice position in the exclusion process (and has no relation to the continuous time of the dynamics of the exclusion process).

Finally, in \cite{Evans2009} it was shown how the trajectories of the queueing system are counted within the matrix product formulation. The  vector $|0\rangle$ corresponds to  an initial queue of length  0. At each discrete time step of the queuing system a matrix $D$ corresponds  to a service time of the queue and a matrix $E$ to a  non-service  time. For example, the action of $D$ on a queue of length $n >0$, represented by $|n\rangle$, is
\begin{equation}
D|n\rangle = |n\rangle + |n-1\rangle\;.
\label{Dac}
\end{equation}
There are two possible events at the service time, which correspond to the two terms in (\ref{Dac}): the first is the service of a new arrival, so that $n$ remains unchanged; the second is  a service and no new arrival, so that $n$ decreases by one. The full details of the queue dynamics are presented in \cite{Evans2009}.

The trajectory of the length of the queue in the queueing process is precisely a Motzkin path, as defined in Section~\ref{sec:DyckMotzkin}. Thus this completes the cycle of mappings from matrix product to multiline configurations to queue trajectories to Motzkin paths for the case $q=0$.

Very recently \cite{Martin2018} a generalised queueing construction, in which each potential service is unused with probability $q^k$ when the queue-length is $k$ was shown to give a recursive construction of the stationary distribution of multispecies process with asymmetry parameter $q$. 

\subsection{Multispecies generalisation}
\label{sec:tensor}

The second-class particle problem is generalised in a straightforward way to the multispecies problem where the class (or species label) of a particle is an integer between $1$ to $N$. Thus there are $N$ classes of particle along with holes, which could be considered as the lowest class of particle. In the partially asymmetric case the dynamics is
\begin{eqnarray}
J\;K \to K\;J \quad\mbox{with rate} \quad 1\quad \mbox{if} \;1\leq J < K \leq N \;,\\
K\;J \to J\;K \quad\mbox{with rate} \quad q \quad \mbox{if} \;1\leq J < K \leq N \;, \\
J\;0 \to 0\;J \qquad\mbox{with rate} \quad 1 \quad \mbox{if} \;1\leq J \leq N \;, 
\\
0\;J \to J\;0 \qquad\mbox{with rate} \quad q
\quad \mbox{if} \;1\leq J \leq N \;.
\end{eqnarray}
The dynamics (\ref{eq:multi}) may also be generalised to have different rates for different particle exchanges---sometimes this is referred to as an `inhomogeneous' multispecies system \cite{AM2013,AL2014}.

The totally asymmetric case ($q=0$) was first constructed by Ferrari and Martin \cite{Ferrari2007} where their queueing interpretation  was generalised to the $N$-species system. This involved the introduction of a system of tandem priority queues with increasing number of classes of customers in each queue. Similarly to the 2-line process corresponding to the 2-TASEP, Ferrari and Martin \cite{Ferrari2007} defined an $N$-line process in which each dynamical event corresponds precisely to a dynamical event in the $N$-TASEP. The steady-state measure of the $N$-line system is just a uniform distribution of particles.  The $N$-TASEP stationary state is found by counting the multiline configurations which map onto particular $N$-TASEP configurations. This combinatorial task is then performed with a matrix product formulation. 

Meanwhile, the matrix product solution for the totally asymmetric case ($q=0$) was constructed in \cite{Evans2009} and hierarchical structure for increasing $N$ was elucidated.  In these solutions  the `matrices' are in fact tensor products of the fundamental matrices $\delta = D- {\bf 1}$, $\epsilon = D- {\bf 1}$ and $A$ and obey more complicated relations than (\ref{DE}--\ref{DA}), involving an auxiliary set of `hat' operators. In \cite{Evans2009} it was shown how the matrices generate the system of priority queues defined in \cite{Ferrari2007}.

To illustrate how the matrix product generalises from the 2-TASEP case to higher numbers of species and what we mean by tensor operators,  let us write down the operators required for the 3-TASEP.
\begin{eqnarray}
  X_1 &=&  {\bf 1}\otimes{\bf 1}\otimes D + 
  \delta \otimes \epsilon \otimes A +  \delta \otimes {\bf 1}\otimes E
\label{MP31} \\
  X_2 &=& A \otimes {\bf 1}\otimes A
    +  A \otimes \delta \otimes E 
\label{MP32} \\
   X_3 &=&  A \otimes A  \otimes E  
\label{MP33} \\
   X_0 &=&   {\bf 1}\otimes{\bf 1}\otimes E +
      {\bf 1}\otimes \epsilon \otimes A + 
  \epsilon \otimes {\bf 1}\otimes D \, .
\label{MP30} 
\end{eqnarray}
These tensor product operators act on state vectors 
\begin{equation}
|l\, m\, n\rangle \equiv |l\rangle \otimes |m\rangle \otimes | n\rangle
\end{equation}
where $|l\rangle =0$ for $l<0$.

It may be shown \cite{Evans2009} how a matrix product using $X_0$, $X_1$, $X_2$, $X_3$ defined in (\ref{MP31}--\ref{MP30}) precisely enumerate the possible trajectories of the state of the tandem queues giving rise to a given configuration $(\tau_1,\tau_2,\dots \tau_N)$.

The matrix product solution was generalised to the partially asymmetric case ($q>0$) in \cite{Prolhac_2009}.
A  class of open boundary multispecies processes has been shown to have a matrix product solution \cite{Finn_2018}.
Further properties of the matrix product formulation and its hierarchical structure have been explored in \cite{Arita2011, Arita2012} for the totally asymmetric and partially symmetric cases. Algebraic properties have been used to generate alternative matrix representations of the Ferrari-Martin construction \cite{Kuniba_2015, Kuniba_2016}. 

Finally we mention  recent work \cite{CGW2015}, in which it was shown how certain Macdonald polynomials, which are a set of orthogonal polynomials with some remarkable properties, may be expressed as a matrix product. The matrix product involves $q$-deformed bosonic operators as does the matrix product for ASEP, which we have discussed in this review. In turn the matrix product formula for Macdonald polynomials can be interpreted in terms of lattice paths leading to a combinatorial interpretation. Also, Macdonald polynomials may be used to express the partition function of the multispecies ASEP. Continuing in this vein a matrix product formula for Koornwinder polynomials has been obtained \cite{CGGW2016} and the multispecies ASEP has been used to derive new combinatorial formulae for Macdonald polynomials \cite{CMW2015}.

\section{Conclusion}\label{sec:conc}

In this work we have reviewed the connection between the stationary weights of configurations in a paradigmatic nonequilibrium statistical mechanical system (the asymmetric simple exclusion process) and combinatorial enumeration problems, such as counting lattice paths. The earliest solutions of the TASEP
(the version of the process in which particles can hop only to the right) appealed to recursion relations \cite{derrida1992exact,schuetz1993phase} between configurational weights which can be expressed more powerfully in terms of reduction relations for matrices \cite{derrida1993exact}, as described in Section~\ref{sec:ASEP}. Both the application of recursion relations and the reordering of matrices implicitly define some kind of counting problem. However it is not necessarily obvious from the outset what is being counted. The generalisation to partially asymmetric hopping resulted in more complicated counting problems involving the parameter $q$.

The most straightforward way to relate the matrix product solution to a lattice path enumeration problem is to exploit a representation of the matrices in terms of the identity, and (in general, $q$-deformed) raising and lowering operators (Section~\ref{sec:explicit}). A particular configuration of the ASEP can then be related to a set of Motzkin paths, in which the identity, raising and lowering operators generate steps that are either horizontal, rise upwards, or fall down. Since the matrices are semi-infinite, the paths may not fall below the origin. Thus one set of objects that are being enumerated by the ASEP normalisation is the set of all paths subject to this constraint. This in turn yields a connection to the Catalan numbers, which solve a large number of enumeration problems \cite{stanley1999enumerative}.

Perhaps one of the most appealing representations of a configurational weight in the TASEP  is in terms of dominated paths (Section~\ref{sec:PathDominanceMapping}). Here, a configuration of the TASEP is converted to a path on the square lattice by drawing (in sequence) a vertical step for each particle and a horizontal step for each empty site (hole). The number of paths that fall below this dominant path, and that have the same start and end point, then gives the weight of the TASEP configuration when $\alpha=\beta=1$. 

Here we see clearly the general phenomenon whereby a configurational weight in the TASEP is given by a sum over a set of objects with simpler weights that live in a larger space. In the specific case of the dominated paths, the larger space is the set of all lattice paths of a fixed length, and the weights are a power of $\alpha$ multiplied by a power of $\beta$. As discussed in Section~\ref{sec:matrixproducts}, we can think of this as a Boltzmann weight, in which $\alpha$ and $\beta$ are the exponential of energetic contributions associated with specific steps along the paths. 

From a practical point of view, the mapping to enumeration problems can expedite the calculation of physical quantities. For example, we saw in Section~\ref{sec:AlphaBeta1TASEP} that once the mapping is established, results from enumerative combinatorics can be used to establish certain quantities more easily than deriving them from scratch via the matrix product solution. In the case of the R\'enyi entropy (Section~\ref{sec:LambdaWeightEnumeration}), it is only by appealing to the lattice walk picture that this has (yet) been calculated, and then only for the case $\lambda=2$. Perhaps there is scope to exploit the mappings further to extend to general $\lambda$, or to obtain new results for systems other than the variants of the asymmetric exclusion process that we have considered here. Finally, it would be of interest to identify cases where generalisations of enumeration problems provide solutions to nonequilibrium stochastic dynamical systems, particularly if these correspond to models for which a matrix product solution has not previously been found.

We have also seen for the open TASEP in Section~\ref{sec:completeconfigs} and for the multispecies TASEP in Section~\ref{sec:multi} that it is possible to construct a Markov process on an   extended configuration space that converges to the Boltzmann-like distribution described above, with the additional feature that the dynamics in the physical region or physical projection is ASEP-like. In the extended space, the dynamics satisfies a dynamic reversibility condition \cite{kelly1979reversibility} (equivalently a bijection in the transition rates---see Appendix~\ref{sec:bijection}), which is essentially a form of detailed balance. The idea that we can have a reversible dynamics in the extended space, but an irreversible dynamics in only part of it, can be understood from an information-theoretic perspective. The dynamics set out in Section~\ref{sec:completeconfigs} induces correlations between the two rows of particles. Projecting onto the subset of physical configurations amounts to erasing these correlations, which in turn causes a loss of information. Since information loss is associated with entropy production, the dynamics in the subsystem can have an irreversible character, even though the dynamics in the full system does not. 

It would be of interest to apply this idea to systems where matrix product solutions exist and try to construct the extended configuration space and dynamics. For example there are several open boundary multispecies models where matrix product solutions exist but for which combinatorial interpretations are not yet clear \cite{Evans1995,Arita2006,Ayyer2009,Crampe2016,Aas2019,Finn_2018}.
An intriguing question  the conditions under which  any irreversible stochastic process might be obtainable from a reversible dynamics on a larger space by projecting onto the physical space.

Most generally, it is well-known that the stationary solution of any Markov process on a finite configuration space can be obtained via the matrix tree theorem \cite{Harary,schnakenberg1976network}. Specifically, one enumerates spanning in-trees on the directed graph of configurations where the edges are weighted by the rate at which a transition between the configurations takes place. In this way, one will {\em always} arrive at a configurational weight that is a sum of products of the elementary transition rates, and where there is a one-to-one correspondence between terms in the sum and spanning in-trees.

In principle, however, the space of trees is very large: since the configurational weight can be expressed as a determinant of an $(n-1)$-dimensional matrix, where $n$ is the number of configurations, we might expect the number of trees to be $\mathcal{O}(\e^N!)$ where $N$ is the number of lattice sites. But in the case of the ASEP, the size of the extended configuration space
(either lattice paths which grow exponentially with $N$ or permutations which grow factorially with $N$) is much smaller than the number of spanning trees implying a massive degeneracy in the weights of spanning trees.  These observations show that it is difficult to predict \emph{a priori} how large the extended space of configurations will be, and furthermore, that its size can be different in different parameter regimes.

\ack{AJW acknowledges studentship funding from the EPSRC under grant number EP/L015110/1.}

\section*{References}
\bibliographystyle{iopart-num}
\bibliography{CombinatorialReview}

\appendix
\renewcommand{\thesubsection}{\Alph{section}.\arabic{subsection}}

\section{Exact expression for ASEP partition function}
\label{sec:Zgen}
Here we present a general expression for the PASEP partition function derived in \cite{blythe2000exact} and its specialisation to the case $\alpha=\beta =1$.
\begin{equation}
\label{eqn:Zsumrep:q<1}
Z_N= \left( \frac{1}{1-q} \right)^{\!\!N}
\sum_{n=0}^N R_{N,n}(q) B_n(v,w;q) \;,
\end{equation}
where
\begin{equation}
\label{eqn:Rdef}
\fl R_{N,n}(q)=\sum_{k=0}^{\halfint{N-n}} (-1)^k \binom{2N}{N-n-2k}
q^{\binom{k}{2}} \left\{ \qbinom{n+k-1}{k-1} + q^k \qbinom{n+k}{k}
\right\}
\end{equation}
which may be alternatively written as
\begin{equation}
\label{eqn:Rdef2}
\fl R_{N,n}(q)=\sum_{k=0}^{\halfint{N-n}} (-1)^k
\left[ \binom{2N}{N-n-2k} - \binom{2N}{N-n-2k-2} \right]
q^{\binom{k+1}{2}} \qbinom{n+k}{k}
\end{equation}
and 
\begin{equation}
\label{eqn:Bdef}
B_n(b,a;q)=\sum_{k=0}^{n} \qbinom{n}{k} b^{n-k} a^k \;.
\end{equation}
In (\ref{eqn:Bdef})
$a$ and $b$ are given, as in the main text, as  \eqref{eq:abdefinition} 
\begin{equation}
a = \frac{1-q-\alpha}{\alpha}\;,\qquad b = \frac{1-q-\beta}{\beta}
\end{equation}
and we have used the $q$-binomial coefficient
\begin{equation}
\label{eqn:qbinomial}
\qbinom{n}{k}= \frac{(q;q)_n}{(q;q)_{n-k} (q;q)_k}\;,
\end{equation}
where the `$q$-shifted factorial' is defined through
\begin{eqnarray}
\label{eqn:qfacdef}
(q;q)_n &=& \prod_{j=1}^{n} (1-q^j)\;, \\
(q;q)_0 &=& 1 \;.
\end{eqnarray}
In the case $\alpha=\beta=1$, $a=b=-q$ and
\begin{equation}
\label{eqn:Bdef}
B_n(b,a;q)=(-q)^n \sum_{k=0}^{n} \qbinom{n}{k} \;.
\end{equation}

\section{Catalan number and Narayana number derivations from matrix representation}
\label{sec:AppendixCatalanDerivation}

Here we first show how the Catalan number result for the $\alpha = \beta = 1$ TASEP partition function \eqref{eq:TASEPPartitionFunctionExpressionab1} is easily obtained from the matrix representation.

Given the $q=0$ ladder operator representation in (\ref{eq:AlphaBeta1LadderRep1}--\ref{eq:AlphaBeta1LadderRep2}), the following recursion relation follows 
\begin{eqnarray}
\fl \bra{n}C^N\ket{m} = & \bra{n-1}C^{N-1}\ket{m} + 2\bra{n}C^{N-1}\ket{m} +\bra{n+1}C^{N-1}\ket{m} 
\end{eqnarray}
with the boundary conditions
\begin{equation}
\bra{-1}C^N\ket{m} = \bra{n}C^N\ket{-1} = 0 \;.
\end{equation}
It is simple to check that the recursion and boundary conditions are satisfied by \cite{derrida1993exact}
\begin{equation}
\bra{n}C^N\ket{m} = \binom{2N}{N+n-m} - \binom{2N}{N+2+n+m}
\end{equation}
and that for $n=m=0$ we obtain for $Z_N$ the Catalan number $C_{N+1}$, using definition \eqref{Z11}
\begin{equation}
Z_N = \bra{0}C^N\ket{0} = \frac{(2N+2)!}{(N+2)!(N+1)!}= C_{N+1}\;.
\label{Zapp}
\end{equation}

We now show how Narayana numbers may be obtained in a similar fashion. Define $G(N,P)$ as the sum of all products of $P$ $D$-matrices and $(N-P)$ $E$-matrices. Using the form of the matrix representation (\ref{eq:AlphaBeta1LadderRep1}--\ref{eq:AlphaBeta1LadderRep2}), the following recursion relation holds \cite{derrida1993exactdiff}
\begin{eqnarray}
\fl \bra{n}G(N,P)\ket{m} =& \bra{n}G(N-1,P-1)\ket{m} + \bra{n}G(N-1,P-1)\ket{m} \\
& + \bra{n+1}G(N-1,P-1)\ket{m} + \bra{n-1}G(N-1,P)\ket{m} \nonumber
\end{eqnarray}
with boundary conditions
\begin{equation}
\bra{-1}G(N,P)\ket{m} = \bra{n}G(N,P)\ket{-1} = 0 \;.
\end{equation}
The recursion and boundary conditions are solved by 
\begin{equation}
\fl \bra{n}G(N,P)\ket{m} = \binom{N}{P}\binom{N}{P+n-m}-\binom{N}{P+1+n}\binom{N}{P-1-m}\;.
\end{equation}
Setting $n = m = 0$, we arrive at the Narayana numbers \eqref{eq:FixedParticleNarayana}
\begin{eqnarray}
\fl \bra{0}G(N,P)\ket{0} &= \binom{N}{P}^2-\binom{N}{P+1}\binom{N}{P-1} \\
& = \frac{N!(N+1)!}{P!(P+1)!(N-P)!(N-P+1)!}
\end{eqnarray}
which is the sum of weights of configurations with $P$ particles.

Summing over all configurations recovers \eqref{Zapp}:
\begin{eqnarray}
Z_N &= \sum_{P=0}^N\left[
\binom{N}{P}^2-\binom{N}{P+1}\binom{N}{P-1}\right] \\
&= \binom{2N}{N} -\binom{2N}{N+2} = C_{N+1}
\end{eqnarray}
where we have used the Vandermonde identity \cite{askey1975orthogonal}
\begin{equation}
\sum_{p=-\infty}^{\infty}\binom{a}{c+p}\binom{b}{d-p} = \binom{a+b}{c+d}\;.
\end{equation}

\section{Demonstration of $DE = D+E$ in path dominance problem}
\label{sec:AppendixReductionDerivation}
Referring to Section~\ref{sec:PathDominanceMapping}, an ASEP configuration $\mathcal{C}$ is uniquely defined by the set of $x$ or $y$-coordinates $\left(x_0, x_1, \dots x_P\right)$, $\left(y_0, y_1, \dots y_{Q}\right)$ that its path $\mathcal{T}$ traces. We now show that the weight of a path $\mathcal{W}(\mathcal{T})$ has the same reduction relation (analogous to $DE = D+E$ \eqref{eq:ReductionRelations1TASEP}), therefore the weight of the TASEP configuration $\mathcal{W}(\mathcal{C}) = \mathcal{W}(\mathcal{T})$.

If we split the path $\mathcal{T}$ into $\mathcal{T} = (\mathcal{T}_{(1)}, \uparrow,\rightarrow, \mathcal{T}_{(2)})$, we are compelled to show relation \eqref{eq:DEPathReduction} (illustrated in Figure \ref{fig:PathReductionRelation2}), which we repeat here:
\begin{equation}
\mathcal{W}(\mathcal{T}) = \mathcal{W}\left(\mathcal{T}_{(1)},\uparrow,\mathcal{T}_{(2)}\right) + \mathcal{W}\left(\mathcal{T}_{(1)},\rightarrow,\mathcal{T}_{(2)}\right)\;.
\end{equation}
This is analogous to the matrix relation $DE = D+E$. If $\mathcal{T}$ is defined by $y$ coordinates $(y_0,y_1,y_2,\dots y_{i-1}, y_i, y_{i+1}, \dots y_{Q-1}, y_Q)$, we therefore require
\begin{eqnarray}
&\mathcal{W}(y_0,y_1,y_2,\dots y_{i-1}, y_i, y_{i+1}, \dots y_{Q-1}, y_Q) \\
= &+\mathcal{W}(y_0,y_1,y_2,\dots y_{i-1}, y_{i+1}, \dots y_{Q-1}, y_Q) \\
& +\mathcal{W}(y_0,y_1,y_2,\dots y_{i-1}, y_i-1, y_{i+1}-1, \dots y_{Q-1}-1, y_Q-1) \;. \nonumber
\end{eqnarray}
We first write $\mathcal{W}\left(\mathcal{T}_{(1)},\rightarrow,\mathcal{T}_{(2)}\right)$ explicitly,
\begin{equation}
\fl \mathcal{W}\left(\mathcal{T}_{(1)},\rightarrow,\mathcal{T}_{(2)}\right) = \sum_{n_0 = 0}^{y_0}\sum_{n_1=n_0}^{y_1} \cdots \sum_{n_{i-1}=n_{i-2}}^{y_{i-1}}\sum_{n_{i}=n_{i-1}}^{y_{i}-1}\sum_{n_{i+1}=n_{i}}^{y_{i+1}-1}\sum_{n_{i+2}=n_{i+1}}^{y_{i+2}-1} \cdots \sum_{n_{Q-1} = n_{Q-2}}^{y_{Q-1}-1} 1
\end{equation}
and rework this expression so to `complete' each of the $n_i, n_{i+1}, \dots n_{Q-1}$ summations sequentially:
\begin{eqnarray}
\fl \mathcal{W}\left(\mathcal{T}_{(1)},\rightarrow,\mathcal{T}_{(2)}\right) = \sum_{n_0 = 0}^{y_0}\sum_{n_1=n_0}^{y_1} \cdots \sum_{n_{i-1}=n_{i-2}}^{y_{i-1}}\sum_{n_i=n_{i-1}}^{y_{i}}\sum_{n_{i+1}=n_{i}}^{y_{i+1}-1}\sum_{n_{i+2}=n_{i+1}}^{y_{i+2}-1} \cdots \sum_{n_{Q-1} = n_{Q-2}}^{y_{Q-1}-1} 1 \\
\fl\quad - \sum_{n_0 = 0}^{y_0}\sum_{n_1=n_0}^{y_1} \cdots \sum_{n_{i-1}=n_{i-2}}^{y_{i-1}}\left(\sum_{n_{i+1}=n_{i}}^{y_{i+1}-1}\sum_{n_{i+2}=n_{i+1}}^{y_{i+2}-1} \cdots \sum_{n_{Q-1} = n_{Q-2}}^{y_{Q-1}-1} 1\right) \nonumber \\
\fl =\sum_{n_0 = 0}^{y_0}\sum_{n_1=n_0}^{y_1} \cdots \sum_{n_{i-1}=n_{i-2}}^{y_{i-1}}\sum_{n_i=n_{i-1}}^{y_{i}}\sum_{n_{i+1}=n_{i}}^{y_{i+1}}\sum_{n_{i+2}=n_{i+1}}^{y_{i+2}-1} \cdots \sum_{n_{Q-1} = n_{Q-2}}^{y_{Q-1}-1} 1 \\
\fl\quad - \sum_{n_0 = 0}^{y_0}\sum_{n_1=n_0}^{y_1} \cdots \sum_{n_{i-1}=n_{i-2}}^{y_{i-1}}\left(\sum_{n_{i+1}=n_{i}}^{y_{i+1}-1}\sum_{n_{i+2}=n_{i+1}}^{y_{i+2}-1} \cdots \sum_{n_{Q-1} = n_{Q-2}}^{y_{Q-1}-1} 1 \right) \nonumber \\
\fl\quad - \sum_{n_0 = 0}^{y_0}\sum_{n_1=n_0}^{y_1} \cdots \sum_{n_{i-1}=n_{i-2}}^{y_{i-1}}\left(\sum_{n_{i+2}=n_{i+1}}^{y_{i+2}-1} \cdots \sum_{n_{Q-1} = n_{Q-2}}^{y_{Q-1}-1} 1 \right) \nonumber \\
\fl =\sum_{n_0 = 0}^{y_0}\sum_{n_1=n_0}^{y_1} \cdots \sum_{n_{i-1}=n_{i-2}}^{y_{i-1}}\sum_{n_i=n_{i-1}}^{y_{i}}\sum_{n_{i+1}=n_{i}}^{y_{i+1}}\sum_{n_{i+2}=n_{i+1}}^{y_{i+2}} \cdots \sum_{n_{Q-1} = n_{Q-2}}^{y_{Q-1}} 1 \\
\fl\quad - \sum_{n_0 = 0}^{y_0}\sum_{n_1=n_0}^{y_1} \cdots \sum_{n_{i-1}=n_{i-2}}^{y_{i-1}}\left(\sum_{n_{i+1}=n_{i}}^{y_{i+1}-1}\sum_{n_{i+2}=n_{i+1}}^{y_{i+2}-1} \cdots \sum_{n_{Q-1} = n_{Q-2}}^{y_{Q-1}-1} 1 \right) \nonumber \\
\fl\quad - \sum_{n_0 = 0}^{y_0}\sum_{n_1=n_0}^{y_1} \cdots \sum_{n_{i-1}=n_{i-2}}^{y_{i-1}}\left(\sum_{n_{i+2}=n_{i+1}}^{y_{i+2}-1} \cdots \sum_{n_{Q-1} = n_{Q-2}}^{y_{Q-1}-1} 1 \right) \nonumber \\
\fl \quad - \dots \nonumber \\
\fl \quad - \sum_{n_0 = 0}^{y_0}\sum_{n_1=n_0}^{y_1} \cdots \sum_{n_{i-1}=n_{i-2}}^{y_{i-1}}\left(\sum_{n_{Q-1}=n_{Q-2}}^{y_{Q-1}-1}1\right) \nonumber \\
\fl \quad - \sum_{n_0 = 0}^{y_0}\sum_{n_1=n_0}^{y_1} \cdots \sum_{n_{i-1}=n_{i-2}}^{y_{i-1}}\left(1\right) \nonumber \;.
\end{eqnarray}
This nested expression then telescopes down to two sums, which can be identified as:
\begin{eqnarray}
\fl \mathcal{W}\left(\mathcal{T}_{(1)},\rightarrow,\mathcal{T}_{(2)}\right) =\sum_{n_0 = 0}^{y_0}\sum_{n_1=n_0}^{y_1} \cdots \sum_{n_{i-1}=n_{i-2}}^{y_{i-1}}\sum_{n_i=n_{i-1}}^{y_{i}}\sum_{n_{i+1}=n_{i}}^{y_{i+1}}\sum_{n_{i+2}=n_{i+1}}^{y_{i+2}} \cdots \sum_{n_{Q-1} = n_{Q-2}}^{y_{Q-1}} 1 \\
\fl\quad - \sum_{n_0 = 0}^{y_0}\sum_{n_1=n_0}^{y_1} \cdots \sum_{n_{i-1}=n_{i-2}}^{y_{i-1}}\sum_{n_{i+1}=n_{i}}^{y_{i+1}}\sum_{n_{i+2}=n_{i+1}}^{y_{i+2}} \cdots \sum_{n_{Q-1} = n_{Q-2}}^{y_{Q-1}} 1 \nonumber
\\
\fl = \mathcal{W}\left(\mathcal{T}\right) - \mathcal{W}\left(\mathcal{T}_{(1)}, \uparrow, \mathcal{T}_{(2)}\right)
\end{eqnarray}
which is the desired result \eqref{eq:DEPathReduction}.

\section{Extended state space TASEP dynamics}
\label{sec:AppendixSimplifiedTwoRow}

\subsection{Bijection implies uniform stationary state}
\label{sec:bijection}
We first note that for a Markov process in which for each transition out
of  a configuration we may associate a transition into the configuration occurring with the same rate, the stationary state has uniform probability.
This is easy to see from the stationarity condition
\begin{eqnarray}
\sum_{\mathcal{C'}} \mathcal{P}^\star(\mathcal{C}) W(\mathcal{C} \to \mathcal{C}') = \sum_{\mathcal{C'}} \mathcal{P}^\star(\mathcal{C}') W(\mathcal{C}' \to \mathcal{C}) \quad \forall \, \mathcal{C}, 
\end{eqnarray}
where $\mathcal{P}^\star(\mathcal{C})$ is the stationary distribution
and $W(\mathcal{C} \to \mathcal{C}') $ is the transition rate from
$\mathcal{C} \to \mathcal{C}'$. 
For $\mathcal{P}^\star(\mathcal{C})$ to be independent of $\mathcal{C}$,
we simply require
\begin{eqnarray}
\sum_{\mathcal{C'}} W(\mathcal{C} \to \mathcal{C}') = \sum_{\mathcal{C'}}  W(\mathcal{C} ' \to \mathcal{C}) \quad \forall \, \mathcal{C}, 
\end{eqnarray}
which is satisfied under the bijection of out-transitions onto in-transitions stated above.

\subsection{Stationarity of simplified two-row TASEP dynamics}
In order to show that the continuous-time Markov process illustrated in Figure~\ref{fig:SimplifiedTwoRow} converges to a stationary distribution that is uniform in the space of complete configurations, we must demonstrate that:
\begin{enumerate}
\item each transition out of a complete configuration is into another complete configuration;
\item every complete configuration can be accessed (by one or more transitions) by any other; and
\item each transition out of a complete configuration can be mapped one-to-one onto a transition into that complete configuration.
\end{enumerate}
Under these conditions the stationary distribution is defined over the full set of complete configurations (and only that set), and is uniform
(see \ref{sec:bijection}).

Recall that the definition of a complete configuration is that the total number of particles on both rows of the lattice up to each site $i=1,2,\ldots N$ must be at least as large as the number of holes. This is, if $n_i$ is the number of particles upto site $i$, we must have $n_i \ge 2i - n_i$, or equivalently, $n_i \ge i$. Note that the zone boundaries drawn in Figure~\ref{fig:SimplifiedTwoRow} indicate the points where $n_i=0$. The two transitions that move particles between the two rows of the lattice leave all $n_i$ unchanged: consequently the final configuration is complete if the initial configuration is complete. If a particle on the bottom row hops to the left, one of the $n_i$ increases by $1$, and the rest remain unchanged. Thus again, an initial configuration that is complete yields to a final configuration that is complete. Conversely, a particle on the top row hopping to the right causes one of the $n_i$ to \emph{decrease} by $1$, and the rest remain unchanged. Thus it could be possible to leave the space of complete configurations if $n_i=0$, i.e., if a particle crosses a zone boundary. Now, if a top-row particle can hop to the right at site $i$, site $(i+1)$ must be empty. The constraint $n_{i+1}\ge0$ implies that we must have a particle in the bottom row at site $(i+1)$, and in fact we must have $n_{i+1}=0$ too. Meanwhile, the constraint $n_{i-1}\ge0$ means that we must have a hole in the bottom row at site $i$. This means that we can arrange for $n_i$ and $n_{i+1}$ to both remain equal to zero if the bottom row particle at $(i+1)$ moves to the left at the same time as the top row particle moves to the right. Since  none of the $n_i$ have changed, this configuration is also complete. This demonstrates  statement (i) above.

To demonstrate statement (ii), we start by identifying two key configurations. The first has the top row empty, and consequently the bottom row full. Given any starting configuration, we can always reach this state through combinations of top-row hops, and movement of the top-right particle onto the bottom row. We can now reach any desired arrangement of particles on the top row by successive movement of the bottom-left particle onto the top row, and top-row hops. During these moves, we have $n_i=0$ for all $i$: this means that every top-row hop from site $i$ to $(i+1)$ will be accompanied by a forced bottom-row hop from $(i+1)$ to $i$. Thus the complete configuration that is arrived at is the one where a hole sits below each top-row particle, and a particle below each top-row hole. There is no complete configuration that has any bottom row particles further to the right than this.  Consequently any desired complete configuration can be reached by moving bottom row particles to the left;  moreover, the top row remains fixed as this is done. Therefore, we have found a path from any complete configuration to any other, which implies that the Markov process is ergodic in the space of complete configurations.

To show that the number of ways into each configuration equals the number of ways out, we need to identify with each transition a `driver' particle. In the cases where only one particle hops, that particle is the driver. These all sit at the front of a domain of particles (in the clockwise hopping direction). In the case where both a top and bottom row particle hop, the top-row particle is the driver (because the bottom row particle is forced to move to maintain positivity). Note that the bottom-row particle is the driver for a different transition. Consequently, the number of ways out of each configuration is equal to the number of domains. This situation with transitions into a configuration is more subtle. Each particle that is at the back of a domain may have moved in the last transition. For particles on the top row, the configuration that has that particle one site to the left (or on the bottom row, if it is the leftmost particle) is also complete, and is therefore one that could have been arrived from. On the bottom row, the configuration that has a particle one site to the right (on on the top row, if it is the rightmost particle) is complete, unless at the target site $i$, $n_i=0$. Movement of the bottom-row particle alone would imply having come from an incomplete configuration, which is not allowed. However, this configuration \emph{can} be reached if a top-row particle in site $(i+1)$ was the driver in the previous move, and forced the bottom-row particle to the left, a move that has not yet been accounted for. Thus to each particle at the back of each domain, we can associate a unique preceding configuration; and consequently the number of ways into each configuration is also equal to the number of domains.

One can also show that these simplified dynamics satisfy a dynamic reversibility condition \cite{kelly1979reversibility}. This involves an involution between complete configurations $\mathcal{C} \to \mathcal{C}^\star$ (i.e., a self-inverse mapping of the set of complete configurations onto itself). Dynamic reversibility is said to hold for some set of transition rates $W(\mathcal{C} \to \mathcal{C}')$ if the following conditions hold:
\begin{eqnarray}
\sum_{\mathcal{C}'} W(\mathcal{C} \to \mathcal{C}') = \sum_{\mathcal{C}'} W(\mathcal{C}^\star \to \mathcal{C}') \quad \forall \, \mathcal{C} \\
\mathcal{P}^\star(\mathcal{C}) W(\mathcal{C} \to \mathcal{C}') = \mathcal{P}^\star(\mathcal{C}'^\star) W(\mathcal{C} '^\star \to \mathcal{C}^\star) \quad \forall \, \mathcal{C}, \mathcal{C}'
\end{eqnarray}
where $\mathcal{P}^\star(\mathcal{C})$ is the stationary distribution. If the `hat' operation is the involution that exchanges the top and bottom rows, the first condition immediately follows from the fact that the number of ways out of each configuration is equal to the number of domains, as this is preserved under the involution. One can check the second condition exhaustively by considering each combination of particles and holes on adjacent sites, and noting that the stationary distribution is uniform.

\section{Demonstration of reduction relations in permutation problem}
\label{sec:AppendixPermutationReductionRelation}

Here we show that the permutation problem detailed in Sections \ref{sec:MappingPermutationProblem}, \ref{sec:qMappingPermutationProblem} has an equivalent reduction relation structure to those in the SSEP and PASEP. Following the formalism in these sections, define
\begin{equation}
\mathcal{W}_{N}\left(\vec{j}\right)
\end{equation}
as shorthand for the total weight of permutations of the integers in $\left(0,1,\dots N\right)$ where only the $\vec{j} = (j_1, j_2, \dots j_P)$ are raised. The reduction relation $\bra{W}E = \bra{W}$ has an equivalent form
\begin{equation}
\label{eq:WEWeqn}
\mathcal{W}_{N+1}\left(\vec{j}+1\right) = \mathcal{W}_{N}\left(\vec{j}\right)
\end{equation}
which is trivial to show: for each permutation on the LHS of \eqref{eq:WEWeqn}, increase every integer by $1$, then append the permutation with a $0$. This gives each permutation on the RHS, with all weights unchanged. Similarly for $D\ket{V} = \ket{V}$,
\begin{equation}
\label{eq:DVVeqn}
\mathcal{W}_{N+1}\left(\vec{j}\right) = \mathcal{W}_{N}\left(\vec{j}\right)
\end{equation}
which can be seen as each permutation on the LHS of \eqref{eq:DVVeqn}, prepended with an $(N+1)$, corresponds to a permutation on the RHS. Again, all weights are unchanged. 

\begin{table}
	\centering
	\begin{tabular}{|lllll|}
		\hline
		Permutation & $\mathcal{W}$ & Consecutive? & New permutation & $\mathcal{W}$ \\
		\hline
		$\underline{0\;4}\;\underline{2\;3}\;1$ & $q^3$ & No & $\underline{1\;4}\;\underline{2\;3}\;0$ & $q^2$ \\
		$\underline{2\;4}\;\underline{0\;3}\;1$ & $q^2$ & No & $\underline{2\;4}\;\underline{1\;3}\;0$ & $q^1$ \\
		$4\;\underline{0\;2\;3}\;1$ & $q^1$ & No & $4\;\underline{1\;2\;3}\;0$ & $q^0$\\
		$\underline{0\;3}\;\underline{2\;4}\;1$ & $q^2$ & No & $\underline{1\;3}\;\underline{2\;4}\;0$ & $q^1$\\
		$\underline{0\;2\;4}\;3\;1$ & $q^2$ & No & $\underline{1\;2\;4}\;3\;0$  & $q^1$\\
		$3\;\underline{0\;2\;4}\;1$ & $q^1$ & No & $4\;\underline{1\;2\;4}\;0$  & $q^0$\\
		$\underline{2\;3}\;\underline{0\;4}\;1$ & $q^1$ & No & $\underline{2\;3}\;\underline{1\;4}\;0$ & $q^0$ \\
		\hline
		$\underline{2\;3}\;1\;\underline{0\;4}$ & $q^0$ & Yes & $\underline{1\;2}\;\underline{0\;3}$ & $q^0$ \\
		$3\;\underline{2\;4}\;\underline{0\;1}$ & $q^0$ & Yes & $2\;\underline{1\;3}\;0$  & $q^0$\\
		$4\;1\;\underline{0\;2\;3}$ & $q^0$ & Yes & $3\;\underline{0\;1\;2}$ & $q^0$\\
		$3\;1\;\underline{0\;2\;4}$ & $q^0$ & Yes & $2\;\underline{0\;1\;3}$  & $q^0$\\
		$\underline{2\;4}\;1\;\underline{0\;3}$ & $q^1$ & Yes & $\underline{1\;3}\;\underline{0\;2}$ & $q^1$ \\
		$1\;\underline{0\;4}\;\underline{2\;3}$ & $q^2$ & Yes & $\underline{0\;3}\;\underline{1\;2}$ & $q^2$\\
		\hline
		$\underline{2\;4}\;3\;\underline{0\;1}$ & $q^1$ & Yes & $\underline{1\;3}\;2\;0$ & $q^1$ \\
		$4\;\underline{2\;3}\;\underline{0\;1}$ & $q^0$ & Yes & $3\;\underline{1\;2}\;0$  & $q^0$\\
		$1\;\underline{0\;3}\;\underline{2\;4}$ & $q^1$ & Yes & $\underline{0\;2}\;1\;3$  & $q^1$\\
		\hline
	\end{tabular}
	\caption{Demonstration of the reduction relation for the configuration $\mathcal{C} = (1010)$, by the reduction $DEDE = qEDDE+DDE+EDE$. $\mathcal{W}(1010)$ is the number of permutations of $(0,1,2,3,4)$ where $0$ and $2$ are raised.}
	\label{tab:PermutationReductionExample}
\end{table}

Finally, the reduction relation $DE = qED + D + E$ has an equivalent form 
\begin{eqnarray}
\fl \mathcal{W}_{N}\left(\vec{j}_1, k, \vec{j}_2\right) \\
 = q\mathcal{W}_{N}\left(\vec{j}_1, k+1, \vec{j}_2\right) + \mathcal{W}_{N-1}\left(\vec{j}_1, k, \vec{j}_2-1\right) + \mathcal{W}_{N-1}\left(\vec{j}_1, \vec{j}_2-1\right) \nonumber
\end{eqnarray}
with all entries of $\vec{j_1}$ less than $k$, and all entries of $\vec{j_2}$ greater than $k+1$. Here, $\mathcal{W}(a,b,c)$ denotes a concatenation of the strings $a$, $b$, $c$.

We prove this by first identifying all LHS permutations where $(\vec{j_1}, k, \vec{j_2})$ are raised (and $k+1$ is not), and $k$, $k+1$ do \emph{not} appear consecutively. We then switch the positions of $k$, $k+1$ in each of these. This then yields all permutations where $(\vec{j_1}, k+1, \vec{j_2})$ are raised, and $k$ is not. 

From the association of weights outlined in Section~\ref{sec:qMappingPermutationProblem}, each of these these new permutations has a weight that is a power of $q$ less than the original permutation. This is the first term of the RHS.

This leaves the permutations on the LHS where $k, k+1$ \emph{do} appear consecutively. If we take these permutations, remove the $k+1$ entry and reduce all integers greater than $k$ by one, we are left with a set of permutations of length $N$ where $(\vec{j_1}, \vec{j_2}-1)$ are raised, and $k$ may or may not be raised. This is the sum of the final two terms of the RHS. See Table \ref{tab:PermutationReductionExample} for a full example of this decomposition, taking configuration $\mathcal{C} = (1,0,1,0)$.

\section{Proof of determinantal form of the partition function}
\label{sec:AppendixDeterminantGeneratingFunction}

We show here that the determinant of the matrix $M$ in \eqref{eq:DeterminantPartitionFunction} is indeed the TASEP partition function, by showing equivalence via generating functions. $M$ is a Hessenberg matrix, which allows its determinant, which we define
\begin{equation}
\mathrm{det}M_{N\times N} \equiv Z'_N
\end{equation} 
to be expressed in a recursive form using \eqref{eq:HessenbergRecurrence}, from Theorem 2.1 in \cite{kaygisiz2013determinants}
\begin{equation}
\label{eq:PartitionFunctionDeterminantRecurrence}
\fl Z'_N = \sum_{r=1}^N(-)^{N-r}Z'_{r-1}\left[\binom{r-1}{N-r-1}\frac{1}{\alpha\beta}+\binom{r-1}{N-r}\left(\frac{1}{\alpha}+\frac{1}{\beta}\right) +\binom{r-1}{N-r+1}\right] \;. \; \;
\end{equation}
We now show that $Z'_N$ and the TASEP partition function $Z_N$ \eqref{eq:TASEPPartitionFunctionExpression} have the same generating function, thus making them equivalent. Define this generating function in $\eta$ as $\mathcal{Z}$
\begin{equation}
\mathcal{Z}(\eta) = \sum_{N\geq0} Z'_N\;\eta^N \;.
\end{equation}
From the recursion \eqref{eq:PartitionFunctionDeterminantRecurrence} and knowing $Z_0 = 1$,
\begin{equation}
\fl \mathcal{Z}(\eta)= 1 + \sum_{N\geq1}\sum_{r=1}^N\eta^N(-)^{N-r}\left[\binom{r-1}{N-r-1}\frac{1}{\alpha\beta}+\binom{r-1}{N-r}\left(\frac{1}{\alpha}+\frac{1}{\beta}\right) +\binom{r-1}{N-r+1}\right]Z'_{r-1} \;. \; \;
\end{equation}
We switch the order of summation and relabel the dummy index $M = N-r$,
\begin{equation}
\fl \mathcal{Z}(\eta) = 1 + \sum_{r\geq1}Z'_{r-1} \eta^{r} \sum_{M\geq0} (-\eta)^{M}\left[\binom{r-1}{M-1}\frac{1}{\alpha\beta}+\binom{r-1}{M}\left(\frac{1}{\alpha}+\frac{1}{\beta}\right) +\binom{r-1}{M+1}\right] \;. \; \;
\end{equation}
Evaluating the summation in $M$,
\begin{equation}
\fl \mathcal{Z}(\eta) = 1+\sum_{r\geq1}Z'_{r-1}\eta^r\left[-\eta(1-\eta)^{r-1}\frac{1}{\alpha\beta}+\left(1-\eta\right)^{r-1}\left(\frac{1}{\alpha}+\frac{1}{\beta}\right)+\frac{1}{\eta}\left(1-(1-\eta)^{r-1}\right)\right] \; \; 
\end{equation}
which we write in terms of the original generating function $\mathcal{Z}$
\begin{eqnarray}
\fl \mathcal{Z}(\eta) = 1+\sum_{r\geq1}Z'_{r-1}(1-\eta)^{r-1}\eta^{r-1}\left[\eta\left(\frac{1}{\alpha}+\frac{1}{\beta}\right)-\frac{\eta^2}{\alpha\beta}-1\right] - \sum_{r\geq1}Z'_{r-1}\eta^{r-1} \\
\fl\;\qquad = \left[\eta\left(\frac{1}{\alpha}+\frac{1}{\beta}\right)-\frac{\eta^2}{\alpha\beta}-1\right]\mathcal{Z}\left(\eta(1-\eta)\right) + \mathcal{Z}(\eta)\;.
\end{eqnarray}
This is factorised to give
\begin{equation}
\fl \mathcal{Z}\left(\eta(1-\eta)\right) = \frac{1}{\left(1-\frac{\eta}{\alpha}\right)\left(1-\frac{\eta}{\beta}\right)} \;.
\end{equation}
Substituting $z = \eta(1-\eta) \Rightarrow \eta = \hf\left(1-\sqrt{1-4z}\right)$,
\begin{eqnarray}
\mathcal{Z}(z) &= \frac{1}{\left(1-\frac{1}{2\alpha}\left[1-\sqrt{1-4z}\right]\right)\left(1-\frac{1}{2\beta}\left[1-\sqrt{1-4z}\right]\right)} \\
& = 1+\left(\frac{1}{\alpha}+\frac{1}{\beta}\right)z + \left(\frac{1}{\alpha}+\frac{1}{\beta}+\frac{1}{\alpha^2}+\frac{1}{\beta^2}+\frac{1}{\alpha\beta}\right)z^2 + \dots 
\end{eqnarray}
which is the known generating function for $Z_N$ (Equation 3.56 in \cite{blythe2007nonequilibrium}). We choose the negative root of $\eta(z)$ to ensure $\mathcal{Z}(0) = 1$.

\end{document}